\documentclass[12pt,preprint]{aastex}
\usepackage{graphicx}
\newcommand{\etal  }{{et al.}}
\newcommand{\msun}{\thinspace M_\odot}  
\newcommand{\rsun}{\thinspace R_\odot}  
\newcommand{\vect}[1]{\mbox{\boldmath$#1$}}

\newcommand{\rhoc}{\rho_{\rm c}}
\newcommand{\nc}{n_{\rm c}}

\newcommand{\cm  }{\,{\rm cm}^{-3} } 

\newcommand{\dfrac}[2]{{\displaystyle \frac{#1}{#2}}  }

\shorttitle{Population III Star Formation}
\shortauthors{Machida  \etal 2006}

\begin{document}
\title{Magneto-Hydrodynamics of Population III Star Formation}

\shortauthors{Machida  \etal 2007}

\author{Masahiro N. Machida\altaffilmark{1}, Tomoaki Matsumoto\altaffilmark{2}, and Shu-ichiro Inutsuka\altaffilmark{1}} 

\altaffiltext{1}{Department of Physics, Graduate School of Science, Kyoto University, Sakyo-ku, Kyoto 606-8502, Japan; machidam@scphys.kyoto-u.ac.jp, inutsuka@tap.scphys.kyoto-u.ac.jp}
\altaffiltext{2}{Faculty of Humanity and Environment, Hosei University, Fujimi, Chiyoda-ku, Tokyo 102-8160, Japan; matsu@i.hosei.ac.jp}

\begin{abstract}
Jet driving and fragmentation process in collapsing primordial cloud  are studied using three-dimensional MHD nested grid simulations.
Starting from a rotating magnetized spherical cloud with the number density of $\nc\simeq10^3\cm$, we follow the evolution of the cloud up to the stellar density $\nc \simeq 10^{22}\cm$. 
We calculate 36 models parameterizing the initial magnetic and rotational energies $(\gamma_0$, $\beta_0)$.
In the collapsing primordial clouds, the cloud evolutions are characterized by the ratio of the initial rotational to magnetic energy, $\gamma_0/\beta_0$.
The Lorentz force significantly affects the cloud evolution when $\gamma_0 > \beta_0$, while the centrifugal force is more dominant than the Lorentz force when $\beta_0 > \gamma_0 $.
When the cloud rotates rapidly with angular velocity of $\Omega_0 > 10^{-17} (\nc/10^3\cm)^{2/3}$\,s$^{-1}$ and $\beta_0>\gamma_0$, fragmentation occurs before the protostar is formed, but no jet appears after the protostar formation.
On the other hand, a strong jet appears after the protostar formation without fragmentation when the initial cloud has the magnetic field of $B_0>10^{-9} (\nc/10^3\cm)^{2/3}$\,G and $\gamma_0 > \beta_0$.
Our results indicate that proto-Population III stars frequently show fragmentation and protostellar jet.
Population III stars are therefore born as binary or multiple stellar systems, and they can drive strong jets, which disturb the interstellar medium significantly, as well as in the present-day star formation, and thus they may induce the formation of next generation stars.
\end{abstract}


\section{INTRODUCTION}
Magnetic field plays an important role in present-day star formation.
Observations indicate that molecular clouds have the magnetic field strengths of order $\sim\mu$G  and magnetic energies comparable to the gravitational energies \citep{crutcher99}.
These strong fields significantly affect star formation processes.
For example, protostellar jets, which are ubiquitous phenomena in the star-forming region, is considered to be driven from the protostar by the Lorentz force \citep{blandford82, pudritz86}. 
The jet influences the gas accretion onto the protostar and disturbs the ambient medium.
In addition, the angular momentum of the cloud is removed by the magnetic braking and protostellar jet.
\citet{tomisaka00} showed that, in his two dimensional MHD calculation, 99.9\% of the angular momentum is transferred from the center of cloud by the magnetic field.
This removal process of angular momentum makes the protostar formation possible in the parent cloud that has much larger specific angular momentum than that of the protostar.
On the other hand, so far, magnetic effects in primordial gas clouds have been ignored in many studies because the magnetic field in the early universe is supposed to be extremely weak.
However, recent studies indicate that moderate strength of magnetic fields exists even in the early universe.
\citet{ichiki06} showed that cosmological fluctuations produce magnetic fields before the epoch of recombination.
These fields are large enough to seed the magnetic fields in galaxies.
\citet{langer03} proposed the generation mechanism of magnetic fields at the epoch of the reionization of the universe.
They found that magnetic fields in the intergalactic matter are amplified up to $\sim 10^{-11}\ {\rm G}$.
These fields can therefore increase up to $\simeq 10^{-7}-10^{-8}\ {\rm G}$ in the first collapsed object having the number density $n \sim 10^3 \cm$.
These fields may influence the evolution of  primordial gas clouds and the formation of Population III stars.

Under the spherical symmetry including the hydrodynamical radiative transfer, the star formation process are carefully investigated by many authors both in the present-day \citep[e.g.,][]{larson69,masunaga00} and primordial star formation \citep[e.g.,][]{omukai98}.
A significant difference between present-day and primordial star formation exists in the thermal evolution of the collapsing gas cloud because of difference in the abundance of dust grains and metals.
In present-day star formation, gas temperature in molecular clouds is $\sim10$\,K.
These clouds collapse isothermally with polytropic index $\gamma \simeq 1$ ($P\propto\rho^{\gamma}$) in $\nc \lesssim 10^{11}\cm$, then the gas becomes adiabatic ($\gamma\simeq 7/5$) at $\nc \simeq 10^{11}\cm$, and an adiabatic core (or the first core) is formed, where $\nc$ denotes the central number density of the collapsing cloud.
When the number density reaches $\nc \simeq 10^{16}\cm$, the molecular hydrogen is dissociated ($\gamma\simeq 1.1$), and the cloud begins to collapse rapidly again.
When the number density reaches $\nc \simeq 10^{20}\cm$, the equation of state becomes hard again ($\gamma \simeq 5/3$). 
The protostar is formed at $\nc \simeq 10^{21}\cm$.
On the other hand, primordial gas clouds have temperature of $\sim200-300$\,K at $\nc \simeq 10^3\cm$ \citep{omukai00,omukai05,bromm02,abel02,yoshida06}.
These clouds collapse keeping $\gamma\simeq 1.1$ for a long range of $10^4\lesssim \nc \lesssim 10^{16}\cm$.
After the central density reaches $\nc \simeq 10^{16}\cm$, the thermal evolution in the primordial collapsing cloud begins to coincide with that in present-day cloud  \citep[for detail, see Fig.~1 of ][]{omukai05}.
The difference in thermal evolution between present-day and primordial clouds arises in the period of $\nc \lesssim 10^{16}\cm$.
%%appears only for $\nc \lesssim 10^{16}\cm$.
In present-day star formation, two distinct flows (molecular outflow and optical jet), which are frequently observed in star forming regions, are driven by the respective core; the outflow is driven by the adiabatic cores (the first core) and the jet is driven by the protostar \citep{tomisaka02,  machida05b, machida07b, banerjee06}. 
In addition, many numerical simulations have shown that fragmentation of the adiabatic core, and thus possible formation of binary or multiple stellar systems.
In contrast, primordial protostars are formed directly without the prior formation of adiabatic core, because the thermal pressure smoothly increases keeping $\gamma \simeq 1.1$ for $\nc\lesssim 10^{16}\cm$.  
Even when the primordial cloud is strongly magnetized, the wide-angle outflow which corresponds to the present-day molecular outflow (that is driven from the adiabatic core formed at $\nc \simeq 10^{11}\cm$) may not appear, but the well-collimated jet (that is driven from the protostar) is ejected at $\nc \gtrsim 10^{21}$.
In addition, it is expected that, in the collapsing primordial cloud, fragmentation rarely occurs for $\nc\lesssim 10^{16}\cm$, because the cloud continues to collapse and the perturbation inducing fragmentation cannot grow sufficiently.

Another major difference between present-day and primordial star formation exists in the magnetic evolution.
In present-day star formation, the neutral gas is coupled well with the ions in $\nc \lesssim 10^{12}\cm$ and $\nc \gtrsim 10^{15}\cm$, while the magnetic field are dissipated by the Ohmic dissipation in $10^{12}\cm \lesssim \nc \lesssim 10^{15}\cm$ \citep[for details, see][]{nakano02}.
\citet{machida07a} showed that $\sim99$\% of the magnetic field in the collapsing cloud is dissipated in $10^{12}\cm \lesssim \nc \lesssim 10^{15}\cm$.
On the other hand, the magnetic evolution in a primordial gas cloud is carefully investigated by \citet{maki04,maki07}, and they found that the magnetic field strongly couples with the primordial gas during all the collapse phase, as long as the initial field strength is weaker than $B_0\lesssim 10^{-5}(n/10^3\cm)^{0.55}$\,G.
\citet{maki07} also showed that ionization fraction is high enough for the magnetic field to couple with the gas even in the accretion phase after the protostar is formed.
In summary, the magnetic field is largely dissipated by the Ohmic dissipation before the protostar formation in present-day cloud, while the magnetic field can continue to be amplified without dissipation in primordial cloud.

Recent cosmological simulations inferred that a single massive star is formed without fragmentation in the first collapsed object that is formed at $\nc \simeq 10^3\cm$ \citep{abel00, abel02,bromm02,yoshida06}.
However, in their simulations, the cloud evolution are investigated only for $\nc \lesssim 10^{17}\cm$.
Since the protostar is formed at $\nc \simeq 10^{21}\cm$ \citep{omukai98}, the protostar is not yet  formed in their simulations.
On the other hand, \citet{machida07b} showed that fragmentation occurs frequently after the equation of state becomes hard for $\nc \gtrsim 10^{17}\cm$, indicating that the binary or multiple stellar system can be formed also in the early universe.

The magnetic field in the collapsing cloud is closely related to fragmentation or formation of binary or multiple stellar systems.  
In present-day star formation, the magnetic field strongly suppresses the rotation driven fragmentation \citep{machida07c,price07}.  
For the primordial cloud, effects of the magnetic field on fragmentation are still unknown. 

In this paper, we investigate the evolution of weakly magnetized primordial clouds and  formation of Population III stars using three-dimensional MHD simulations, and show the driving condition of a jet from proto-Population III stars and the fragmentation condition in the collapsing primordial cloud cores.
The structure of the paper is as follows.
The framework of our models and the numerical method are given in \S 2.
The numerical results are presented in \S 3.  
We discuss the fragmentation and jet driving conditions in \S 4, and summarize our results in \S 5.

\section{MODEL}
 Our initial settings are almost the same as those of \citet{machida06d,machida07d}.
To study the cloud evolution, we use the three-dimensional ideal MHD nested grid code. 
We solve the MHD equations including the self-gravity:  
\begin{eqnarray} 
& \dfrac{\partial \rho}{\partial t}  + \nabla \cdot (\rho \vect{v}) = 0, & \\
& \rho \dfrac{\partial \vect{v}}{\partial t} 
    + \rho(\vect{v} \cdot \nabla)\vect{v} =
    - \nabla P - \dfrac{1}{4 \pi} \vect{B} \times (\nabla \times \vect{B})
    - \rho \nabla \phi, & 
\label{eq:eom} \\ 
& \dfrac{\partial \vect{B}}{\partial t} = 
   \nabla \times (\vect{v} \times \vect{B}), & 
\label{eq:reg}\\
& \nabla^2 \phi = 4 \pi G \rho, &
\end{eqnarray}
 where $\rho$, $\vect{v}$, $P$, $\vect{B} $, and $\phi$ denote the density, 
velocity, pressure, magnetic flux density,  and gravitational potential, respectively. 
For gas pressure, we use the barotropic relation that approximates the result of \citet{omukai05}.
They consider detailed thermal and chemical processes in collapsing primordial gas, adopting the a simple one-zone model where the core collapses approximately at the free-fall timescale and the size of the core is about the Jeans length as in the Larson-Penston self-similar solution \citep{penston69,larson69}.
We fit to the thermal evolution derived by \citet{omukai05} as a function of the density \citep[see Fig.1 of][]{machida07d}, and the pressure is calculated using the fitted function.
Therefore, our equation of state is a barotropic.

As the initial conditions of the clouds, we consider the density profile of the critical Bonnor-Ebert sphere \citep{ebert55,bonnor56} with perturbations,
\begin{eqnarray}
\rho(r, \varphi) = \left\{
\begin{array}{ll}
\rho_{\rm BE}(r)\,f \, \left[1+\delta \rho(r, \varphi)\right] & \mbox{for} \; \; r < R_{0}, \\
\rho_{\rm BE}(R_0) \,f\, \left[1+\delta \rho(r, \varphi)\right] & \mbox{for}\; \;  r \ge R_{0}, \\
\end{array}
\right. 
\label{eq:init}
\end{eqnarray}
where $\rho_{\rm BE}(r)$ denotes the density distribution of the critical Bonnor-Ebert sphere.
The central number density of the Bonnor-Ebert sphere is set at $\rho_{\rm BE}(0) = 10^3\cm$.
The factor $f$ denotes a density enhancement factor, and the density is increased by a factor $f =1.86 $ to promote the collapse.
The initial cloud has therefore the central density of $n_0 = 1.86\times10^3\cm$.
The initial temperature is set at 250\,K, and the radius of the Bonnor-Ebert sphere is $R_0 =6.5$\,pc, corresponding to the non-dimensional radius of Bonnor-Ebert sphere, $\xi_{\rm max} = 6.5$. 
Outside this radius, a uniform gas density of $n_{\rm BE}(R_0)=130  \cm$ is assumed.
To promote fragmentation, a small $m=2$-mode density perturbation is imposed to the spherical cloud such as,
\begin{equation}
\delta \rho = A_{\varphi} (r/R_0)\, {\rm cos}\, 2\varphi, 
\end{equation}
where $A_{\varphi}$ is the amplitude of the perturbation and $A_{\varphi}$ = 0.01 is adopted in all models.
The total mass contained inside $R_0$ is $ M_{\rm c} = 1.8\times 10^4 \msun$.
The density enhancement factor $f$ specifies the ratio of the thermal to the gravitational energy $\alpha_0$ ($= E_{\rm th}/|E_{\rm grav}|)$, where $E_{\rm th}$ and $E_{\rm grav}$ are the thermal and gravitational energies of the initial cloud).
The initial cloud has  $\alpha_0 = 0.5$ when $f=1.86$ and $A_\varphi=0$.

The initial cloud rotates around the $z$-axis at a uniform angular velocity of $\Omega_0$,  and it has a uniform magnetic field $B_0$ parallel to the $z$-axis (or rotation axis).
The initial model is characterized by two non-dimensional parameters: the ratio of the rotational energy to the gravitational energy $\beta_0$ ($=E_{\rm rot}/|E_{\rm grav}|$), and the magnetic energy to the gravitational energy $\gamma_0$ ($=E_{\rm mag}/|E_{\rm grav}|$) , where $E_{\rm rot}$ and $E_{\rm mag}$ are the rotational and magnetic energies.

We made 36 models changing these two parameters.
All the models examined here and their parameters ($\beta_0$ and $\gamma_0$) are summarized in Table~\ref{table}.
These tabulated values are calculated in the cases of spherical clouds, i.e., $A_\varphi = 0$.
We examine a large parameter range of $\beta_0=10^{-6}-0.1$, $\gamma_0 = 0-0.2$.
For convenience, $\beta_0$ is referred as ``initial rotational energy", and $\gamma_0$ as ``initial magnetic energy.''
The angular velocity $\Omega_0$ and magnetic field $B_0$ are also summarized in Table~\ref{table}.
In addition, we estimate the mass-to-flux ratio,
\begin{equation}
\dfrac{M}{\Phi} = \frac{M}{\pi R_0^2 B_0},
\end{equation}
where $M$ and $\Phi$ denote the mass contained within the cloud radius $R_0$, and the magnetic flux threading the cloud, respectively.
There exists a critical value of $M/\Phi$ below which a cloud is supported against the gravity by the magnetic field.
For a cloud with an uniform density, \citet{mouschovias76} derived a critical mass-to-flux ratio,
\begin{equation}
\left(\dfrac{M}{\Phi}\right)_{\rm cri} = \dfrac{\zeta}{3\pi}\left(\dfrac{5}{G}\right)^{1/2},
\end{equation}
where $\zeta=0.53$ for uniform spheres \citep[see also,][]{maclow04}.
For convenience, we use the mass-to-flux ratio normalized by the critical value,
\begin{equation}
\left( \dfrac{M}{\Phi}\right)_{\rm norm} \equiv \left(\dfrac{M}{\Phi}\right)/\left(\dfrac{M}{\Phi}\right)_{\rm cri},
\end{equation}
which is summarized in Table~\ref{table}.

We adopt the nested grid method \citep[for details, see ][]{machida05a,machida06a} to obtain high spatial resolution near the center.
Each level of a rectangular grid has the same number of cells ($ = 128 \times 128 \times 64 $), with the cell width $h(l)$ depending on the grid level $l$.
The cell width is halved with every increment of the grid level.
The highest level of grids changes dynamically: a new finer grid is generated whenever the minimum local Jeans length $\lambda _{\rm J}$ falls below $8\, h (l_{\rm max})$, where $h$ is the cell width. 
The maximum level of grids is restricted to $l_{\rm max} = 30$.
Since the density is highest in the finest grid,  generation of a new grid ensures the Jeans condition of \citet{truelove97} with a margin of a safety factor 2.
We begin our calculations with three grid levels ($l=1-3$).
Box size of the initial finest grid $l=3$ is chosen to be $2 R_0$, where $R_0$ is the radius of the critical Bonnor-Ebert sphere. 
The coarsest grid ($l=1$) then has box size of $2^3\, R_0$. 
The mirror symmetry with respect to $z$=0 is imposed.
A boundary condition is imposed at $r=2^3\, R_0$, where the magnetic field and ambient gas rotate at an angular velocity of $\Omega_0$ (for detail, see \citealt{matsu04}).
We adopted the hyperbolic $\nabla \cdot \vect{B}$ cleaning method of \citet{dedner02}.

\section{RESULTS}
\subsection{Typical Jet Model}
\label{sec:out}
\begin{figure}
\begin{center}
\includegraphics[width=160mm]{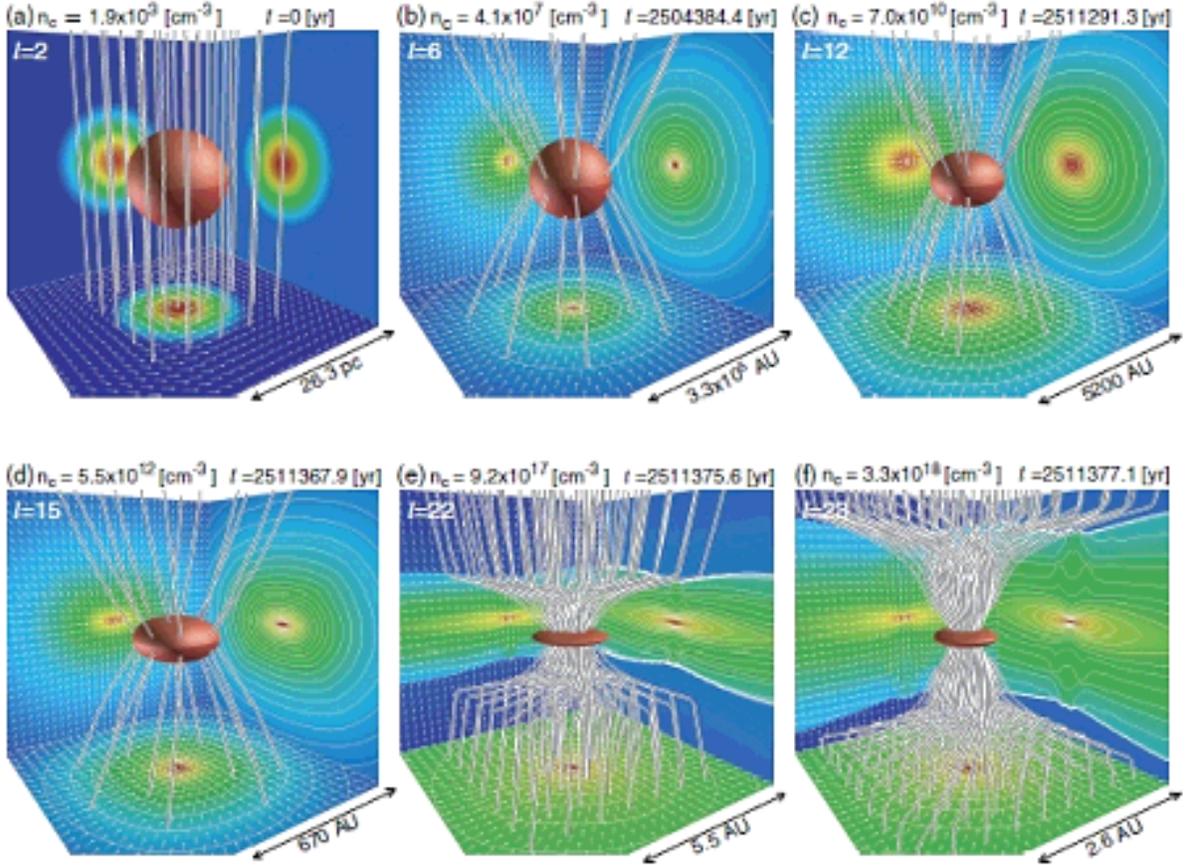}
\caption{
({\it a}--{\it f}) Time sequence of model 23 [($\beta_0$, $\gamma_0$) = ($10^{-3}$, $2\times10^{-3}$)] before the protostar formation $\nc < 10^{21}\cm$.
In each panel, the structures of the high-density region ({\it red isosurface}) and magnetic field lines ({\it black and white streamlines}) are plotted in three dimensions, while the density contours ({\it false color} and {\it contour lines}) and velocity vectors ({\it thin arrows}) are projected in each wall surface.
The central number density $\nc$, elapsed time $t$, grid level $l$, and grid size are also shown in each panel.
}
\label{fig:1}
\end{center}
\end{figure}

Firstly, we show a cloud evolution with a clear jet from a non-fragmentation core.
Figure~\ref{fig:1} shows the cloud evolution before the protostar formation ($\nc<10^{21}\cm$) from the initial stage  for model 23.
Model 23 has parameters of ($\gamma_0$, $\beta_0$) = ($2\times10^{-3}$, $10^{-4}$), and the initial cloud has a magnetic field strength of $B_0 = 10^{-6}$\,G, and a angular velocity of $\Omega_0=2.3\times10^{-16}$\,s$^{-1}$.
Figure~\ref{fig:1}{\it a} shows the initial spherical cloud (i.e., Bonnor-Ebert Sphere) threaded by the uniform magnetic field.
Figures~\ref{fig:1}{\it b}--\ref{fig:1}{\it d} show the cloud structure around the center of cloud when the central density reaches $\nc=$ ({\it b}) $4.1\times10^7\cm$, ({\it c}) $7.0\times 10^{10}\cm$, and ({\it d}) $5.5\times10^{12}\cm$, respectively.
The density contours projected on the sidewall in these figures indicate that the central region becomes oblate as the cloud collapses because of magnetic field and rotation, both of which are amplified by the cloud collapse.
Figures~\ref{fig:1}{\it a}--\ref{fig:1}{\it d} also show that the magnetic field lines are gradually converged toward the center as the central density increases.

Figure~\ref{fig:1}{\it e} shows the cloud structure at $\nc = 9.2\times 10^{17}\cm$, where a thin disk is formed at the center of cloud.
In this figure, the disk-like structure is threaded by the magnetic field lines that are strongly converged toward the center of the cloud.
The density contours projected on the sidewall of Figure~\ref{fig:1}{\it e} show the shock waves above and below the disk (see the crowded contour lines on the right sidewall).
Near the shock surface, the magnetic field lines are bent. 
Similar structures are seen in the present-day star formation process \citep{machida05b,machida06a}.
We discuss the relation between the shock generation and the evolution of the magnetic field and rotation in \S\ref{sec:dis-convergence}.
The cloud structure at $\nc = 3.3\times 10^{18}\cm$ are shown in Figure~\ref{fig:1}{\it f}, in which the magnetic field lines are strongly converged but slightly twisted.
Also in this figure, clear shocks are seen above and below the disk-like structure.
Figures~\ref{fig:1}{\it a}--\ref{fig:1}{\it f} show that the magnetic field lines are hardly twisted before the protostar formation ($\nc \lesssim 10^{21}\cm$), because the collapse timescale is shorter than the rotational timescale.

\begin{figure}
\begin{center}
\includegraphics[width=160mm]{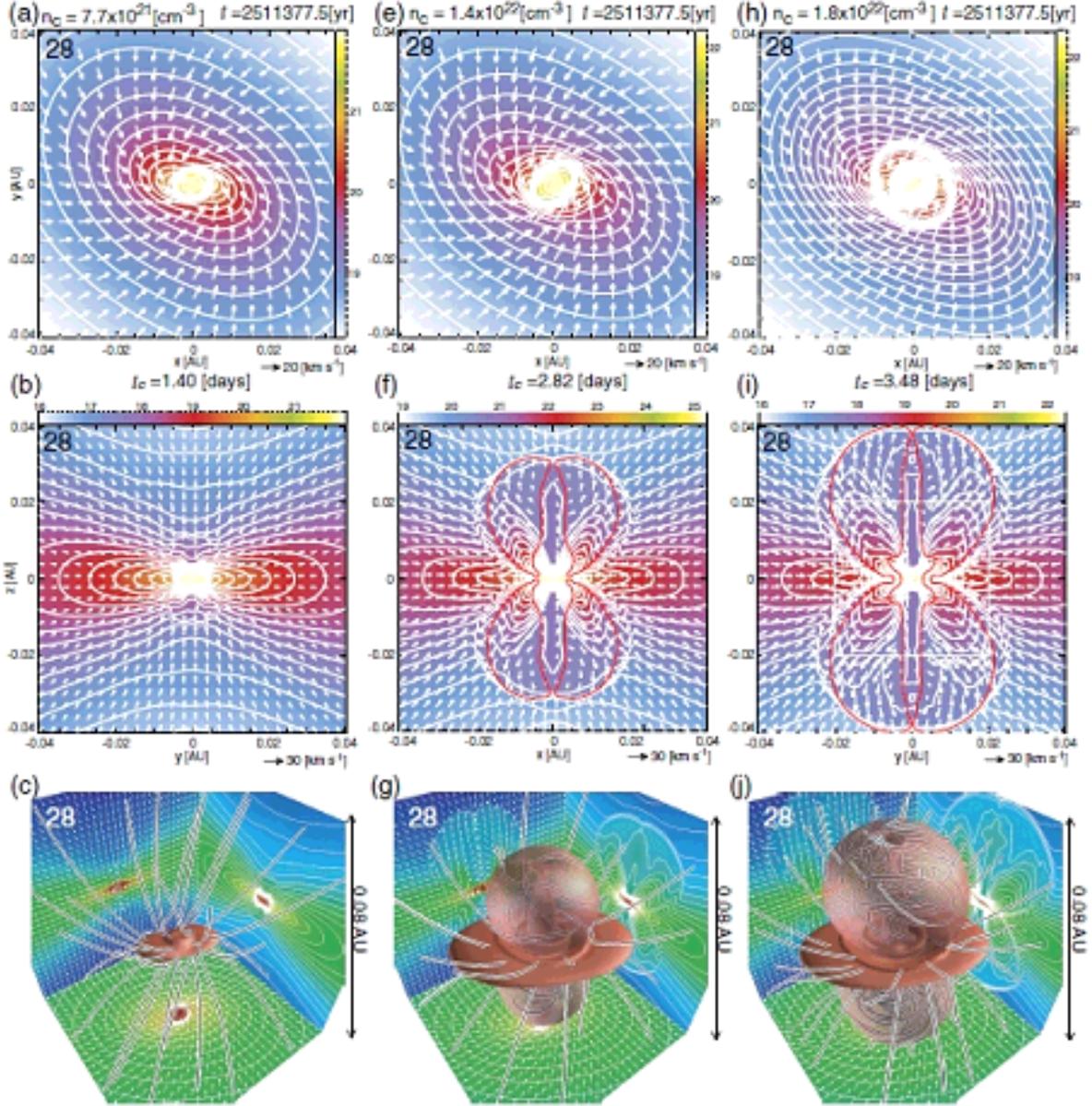}
\caption{
Time sequence of model 23 [($\beta_0$, $\gamma_0$) = ($10^{-3}$, $2\times10^{-3}$) ] with $l=28$ level of grid after the protostar formation $\nc > 10^{21}\cm$.
The structure around the center of the cloud is plotted on $z=0$  plane (upper panels), and $y=0$ plane (middle panels), and in three dimension (lower panels).
The density distribution ({\it false color} and {\it contours}) and velocity vectors ({\it arrows}) are plotted in the upper and middle panels.
The red curves denote a border between inflow and outflow regions in the upper panels.
The magnetic field lines ({\it black and white streamlines}), high-density region ({\it red isosurface}), and outflow region ({\it transparent red isosurface}) are plotted in each lower panel.
The density contours ({\it false color} and {\it contour lines}) and velocity vectors ({\it thin arrows}) are also projected on each wall surface in each lower panel.
The central number density $\nc$, and elapsed time $t$ are denoted at the top of each upper panel.
The elapsed time after the protostar formation $t_{\rm c}$ is shown at the bottom of each upper panel.
}
\label{fig:2}
\end{center}
\end{figure}

Figure~\ref{fig:2} shows the cloud evolution after the protostar formation ($\nc>10^{21}\cm$).
The upper and middle panels show the density and velocity distributions on $z=0$ and $x=0$ plane, respectively, and the lower panels show the cloud structures and configurations of the magnetic field lines in three dimensions.
Figures~\ref{fig:2}{\it a}--\ref{fig:2}{\it c} show the cloud structure $1.4$\,days after the protostar formation.
In this model, when the central density reaches $\nc\simeq 4.7\times 10^{21}\cm$, a protostar surrounded by the strong shock is formed around the center of the cloud.
The protostar has a mass of $M=7.6\times10^{-3}\msun$, and a radius of $r=1.1\rsun$ at its formation epoch.
The shock front corresponding to protostellar surface is seen in Figure~\ref{fig:2}{\it a} ($r\sim0.005$\,AU; contours of $n\sim10^{21}\cm$).
The density contours projected on the sidewall in Figures~\ref{fig:2}{\it c} also show the shock surface near the center of the cloud.
The magnetic field lines are strongly converged to the center of the cloud, but slightly twisted at this epoch.
The jet starts to be ejected $1.68$\,days after the protostar formation.
Figures~\ref{fig:2}{\it e}--\ref{fig:2}{\it g} shows the cloud structure $2.82$\,days after the protostar formation.
The jet ejected near the protostar is indicated by the red contour in Figure~\ref{fig:2}{\it f} and the red transparent surface in Figure~\ref{fig:2}{\it g}, which are a boundary between the inflow ($v_r < 0$) and jet ($v_r > 0$) regions.
The magnetic field lines are twisted significantly inside the jet region because of the rapid rotation of the protostar.
The jet affects the density distribution;
the butterfly-like density distribution is shown in Figures~\ref{fig:2}{\it f} and \ref{fig:2}{\it g}  (see contour lines near the protostar projected on the right sidewall).
The strong shock waves are generated at the upper and lower ends of the jet, reflecting a strong mass ejection.

Figures~\ref{fig:2}{\it h}--\ref{fig:2}{\it j} show the cloud structure 3.48\,days after the protostar formation, the last stage of this calculation.
Until the last stage, the jet continue to extend as is seen in Figures~\ref{fig:2}{\it f} and \ref{fig:2}{\it i}.
In Figure~\ref{fig:2}{\it h}, two nested shocks are observed near the center of the cloud ($r\sim0.005$\,AU and $r\sim0.01$\,AU).
The outer shock is generated by the mass ejection (i.e., jet) near the protostar, while the inner one corresponds to the surface of the protostar.
In Figure~\ref{fig:2}{\it j}, the isodensity surface exhibits a shallow cone-shape on the disk, which corresponds to the outer shock in Figure~\ref{fig:2}{\it h}.
The magnetic field lines inside the jet region are more strongly twisted with times as shown in Figures~\ref{fig:2}{\it g} and \ref{fig:2}{\it j}.
At the end of the calculation, the jet has a maximum speed of $v_{\rm max}=66.3$\,km, and it extends up to  0.05\,AU.
The mass of protostar reaches  $M_{\rm ps}= 8.5\times 10^{-3}\msun$, while mass of the outflowing gas is $M_{\rm out}= 1.1\times 10^{-3}\msun$.
About 10\% of the accreting matter is therefore ejected from the center of the cloud via the protostellar jet.
The mass ejection rate is estimated as $\dot{M} = 5.9\times10^{-3}\msun$\,yr$^{-1}$.

\subsection{Typical Fragmentation Model}
\begin{figure}
\begin{center}
\includegraphics[width=160mm]{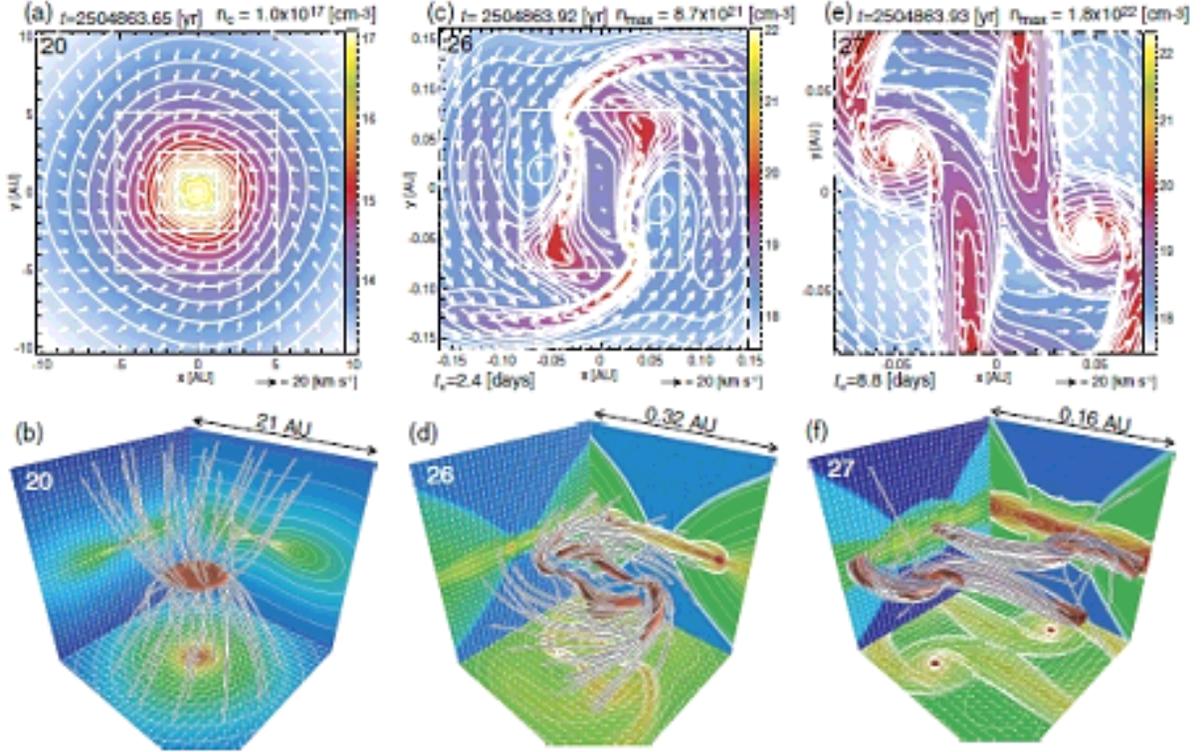}
\caption{
Time sequence of model 21 [($\beta_0$, $\gamma_0$) = ($10^{-4}$, $2\times10^{-7}$) ].
The density distribution ({\it false color} and {\it contours}) and velocity vectors ({\it arrows}) on the $z=0$ plane are plotted in each upper panel.
The magnetic field lines ({\it black and white streamlines}) and high-density region ({\it red isosurface}) in three dimension  are plotted in each lower panel.
In each lower panel, the density contours ({\it false color} and {\it contour lines}) and velocity vectors ({\it thin arrows}) are also projected on each wall surface.
The central number density $\nc$, elapsed time $t$, and elapsed time after the protostar formation $t_{\rm c}$ are denoted in each upper panel.
The grid level is shown in the upper left corner of each panel.
}
\label{fig:3}
\end{center}
\end{figure}

Next, we describe the cloud evolution exhibiting fragmentation.
Figure~\ref{fig:3} shows the cloud evolution for model 21 with 
($\gamma_0$, $\beta_0$) = ($2.5\times10^{-7}$, $10^{-4}$), which are corresponding to a magnetic field strength of $B_0 = 10^{-8}$\,G and an angular velocity of $\Omega_0=2.3\times10^{-16}$\,s$^{-1}$ at the initial stage.
Model 21 has the same rotational energy as the model 23, the previous model, but $10^{-4}$ times smaller magnetic energy than model 23.
Figures~\ref{fig:3}{\it a} and \ref{fig:3}{\it b} show the cloud structure just before the protostar formation ($n_c=1.0\times 10^{17}\cm$).
We define the protostar formation epoch as the stage of $n_c = 10^{21}\cm$.
Similar to model 23, a thin disk is formed at the center of the cloud before the protostar formation.
The disk keeps an almost axisymmetric structure in the plane perpendicular to the rotation axis.
Although we added 1\% of the non-axisymmetric density perturbation to the initial state, the non-axisymmetric structure hardly evolves before the protostar formation.
Figure~\ref{fig:3}{\it b} also shows an hourglass configuration of the magnetic field lines, similar to model 21.

Figures~\ref{fig:3}{\it c} and \ref{fig:3}{\it d} show the cloud structure 2.4\,days after the protostar formation.
The central region is deformed from a disk structure into a ring at $\nc\simeq 8\times 10^{18}\cm$, and the ring fragments.
This is consistent with the fragmentation criterion of \citet{machida07d}; a non-magnetized  cloud with rotational energy of $\beta_0 > 10^{-5}$ undergoes fragmentation. The cloud here has a rotational energy of $\beta_0=10^{-4}$ and the very small magnetic energy of $\gamma_0 = 2.5\times10^{-7}$ at the initial stage.
Fragments are located at ($x$, $y$) $\simeq$ ($\pm$0.035\,AU, $\mp$0.055\,AU) with long spiral tails that are the remnant of the ring (Fig.~\ref{fig:3}{\it c}).
The magnetic field lines are distributed strongly along the spiral tails (Fig.~\ref{fig:3}{\it d}).

Figures~\ref{fig:3}{\it e} and \ref{fig:3}{\it f} show the cloud structure 8.8\,days after the protostar formation.
The fragments exhibit a round shape in Figure~\ref{fig:3}{\it e}.
The central density of the fragments reaches $\nc \simeq 1.8\times 10^{22}\cm$ at this epoch.
Fragments are surrounded by their respective disk with spiral arms, and the disks grow with time, comparing Figures~\ref{fig:3}{\it c} and \ref{fig:3}{\it e}.
At the end of the calculation, each fragment (i.e., each protostar) has a mass of $\sim1.3\times10^{-2}\msun$ and a radius of $\sim1.4\rsun$.
The separation between fragments is $\sim0.89$\,AU.
As shown in Figure~\ref{fig:3}{\it f}, the magnetic field lines are distributed along the spiral tails, indicating that the vertical component of the magnetic field is weaker than the other component (i.e., $B_z \ll B_x, B_y$).  
This is responsible to the weak magnetic field around the protostars, and the magnetic field lines easily follow the orbital motion of the fragments.

This fragmentation model does not exhibit a jet formation by the end of the calculation in spite of the same initial rotation speed as in the previous jet model. 
This is also responsible to the weak magnetic field.
Note that the calculation halts 8.8 10~days after the stellar core formation, and we does not reject a possibility that a jet appears in  later stage.

One may image that this model undergoes the rotation driven fragmentation because of a weak magnetic braking, and the previous model is kept away from the fragmentation because of the strong magnetic braking.  
However, both the models show the same amplification rate for the magnetic field and the angular velocity during the collapse.
The saturation levels of the amplification for the magnetic field and angular velocity depend on the initial conditions.
The two models track two different evolutional sequences according to their respective $\gamma_0/\beta_0$. 
This point is discussed in \S\ref{sec:dis-frag-jet-cond} and \S\ref{sec:summary}.

\subsection{Fragmentation and Magnetic/Rotational Energy}

\begin{figure}
\begin{center}
\includegraphics[width=160mm]{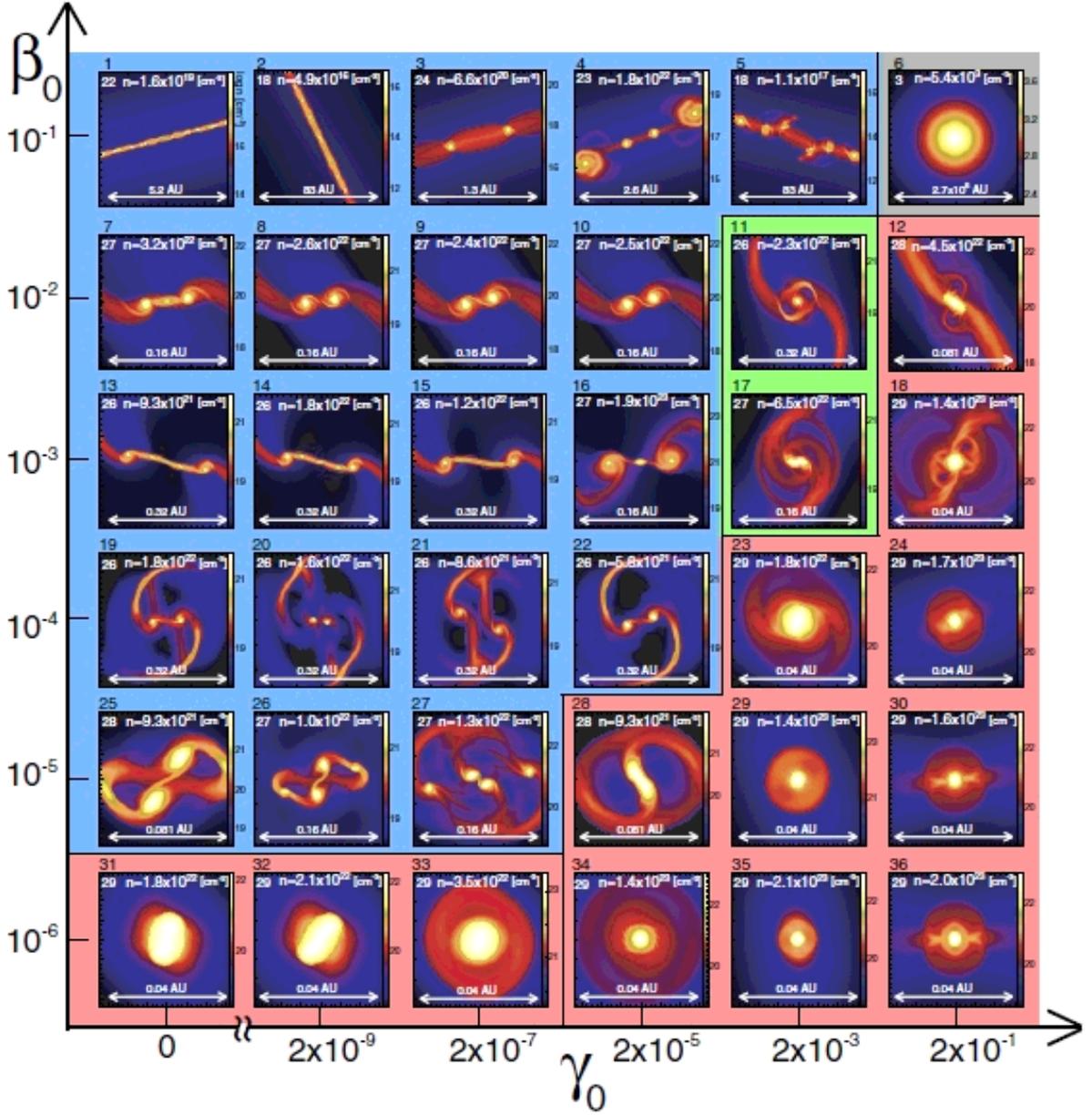}
\caption{
Final states on the $z=0$ plane against parameters $\gamma_0$ and $\beta_0$.
The model numbers are described in the upper left corner outside each panel.
The density distribution (color-scale) is plotted in each panel.
The grid level, maximum number density ($n$), and grid scale are denoted inside each panel.
$\gamma_0$ - $\beta_0$ plane is divided into four regions indicated by colors: the fragmentation ({\it blue}),  non-fragmentation ({\it red}), merger ({\it green}), and  no-collapse ({\it gray}) models.
}
\label{fig:4}
\end{center}
\end{figure}

Figure~\ref{fig:4} shows the final stages in the plane of initial magnetic ($\gamma_0$, $x$-axis) and rotational energies ($\beta_0$, $y$-axis) for every models.
The density distribution in the $z=0$ plane is shown in each panel.
Models are classified into four types: fragmentation models (models in the blue background), non-fragmentation models (models in the red background), merger models in which the fragments merge after fragmentation (models in the green background), and no collapsing model (model in the gray background).
In fragmentation models, calculations were stopped when the Jeans condition was violated in a grid except for the finest grid.
Such a case occurs either when fragments escape from the finest grid or when the gas far from the center becomes denser than that in the finest grid.
In non-fragment models, the Jeans condition was not violated  because the densest gas is located at the finest grid, and the calculations was stopped at some high density.
%%because of the computational cost. 
After the protostar formation, the Alfv\'en velocity becomes extremely large due to the strong magnetic field at the central region, and the timestep becomes extremely short.
We stopped calculation after we checked that fragmentation was not likely to occur around the protostar.
The fragmentation reproduced here is therefore restricted to the cases where fragmentation occurs near the central region and fragmentation occurs just after the protostar formation. 
In merger models [($\gamma_0$, $\beta_0$) = ($2\times10^{-3}$, $10^{-2}$), ($2\times10^{-3}$, $10^{-3}$)],  fragments merge to form a single core at the center of the cloud after fragmentation.
The merged core did not undergo fragmentation again although we calculated merger models for sufficiently long term. 
In the no collapsing model [($\gamma_0$, $\beta_0$) = (0.2, 0.1)], the cloud oscillates without collapse because this model has the large magnetic and rotational energies.

In Figure~\ref{fig:4}, fragmentation models are distributed in the upper left region, indicating that a cloud tends to fragment when the initial cloud has a weaker magnetic field and faster rotation.
Namely, the magnetic field suppresses fragmentation, while the rotation promotes fragmentation.
This effect of the magnetic field and rotation on fragmentation is similar to that in present-day star formation \citep{hosking04,machida04,machida05b,machida07c,price07,hennebelle07a}.
Firstly, to investigate the effect of rotation on fragmentation, we focus on models displayed in the fourth column in Figure~\ref{fig:4} (models 4, 10, 16, 22, 28, and 34).
These models have the same magnetic energies of $\gamma_0 = 2\times 10^{-5}$ and different rotational energies of $\beta_0 = 10^{-6}-0.1$ at the initial stage.
Fragmentation occurs in models with $\beta_0 > 10^{-4}$ (models 4, 10, 16, 22), while fragmentation does not occur in models with $\beta_0 < 10^{-4}$ (models 28, 34).
In addition,  models with larger rotational energies undergo fragmentation in the earlier evolution phases with wider separation.
For example, model 4 ($\beta_0 = 0.1$) fragments at $\nc = 9.5\times10^{17}\cm$ with the separation of $R_{\rm sep} \simeq 4.3$\,AU, while model 22 ($\beta_0=10^{-4}$) does at $\nc = 1.6\times10^{19}\cm$ with the separation of $R_{\rm sep} \simeq 0.2$\,AU.
\citet{machida07d} also showed that a cloud with faster rotation exhibits wider separation.
On the other hand, paying attention to the non-fragmentation models, 
model 34 exhibits a compact spherical protostar with a disk, 
while model 28 shows bar and ring structures.
Model 34 did not show any sign of fragmentation, and the central core continues to grow.
Model 28 has marginal parameters for fragmentation since it is located at the boundary between fragmentation and non-fragmentation models.
The protostar fails to fragment in spite of its bar shape.
The spiral arms evolve into a ring, which does not also fragment.
%%If we adopt different amplitude of initial non-axisymmetric perturbation for model 28, fragmentation may occur.

We now focus on models displayed in the fourth low in Figure~\ref{fig:4} (models 19-24), to investigate the effect of magnetic field on fragmentation.
These models have different magnetic energies of $\gamma_0 = 0-0.2$ but the same rotational energies of $\beta_0 = 10^{-4}$ at the initial stage.
In these models, weakly magnetized clouds undergo fragmentation ($\gamma_0 < 2\times 10^{-5}$, models 19-22), while strongly magnetized clouds do not ($\gamma_0 > 2\times 10^{-5}$, models 23 and 24). 
When clouds have the same rotational energies at the initial stage, model with weaker magnetic field undergo fragmentation at the earlier evolutionary phases with wider separation.
For example, the model with weaker magnetic field $\gamma_0=2\times 10^{-9}$ (model 28) fragments at $\nc =7.8\times 10^{18}\cm$, while the model with stronger magnetic field  $\gamma_0=2\times 10^{-5}$ (model 22) does at $\nc =1.6\times 10^{19}\cm$.
Moreover, the models with much larger magnetic energies (models 23 and 24) produce a single compact core at the center of the cloud.

Fragmentation and non-fragmentation models are clearly separated in Figure~\ref{fig:4}.
In addition, merger models are located at the boundary between fragmentation and non-fragmentation models.
In fragmentation models, models with weaker magnetic field and faster rotation tend to have wider separation.
In non-fragmentation models, clouds have more compact cores in models with stronger magnetic fields and slower rotations.
These features indicate clearly that the rotation promotes fragmentation, and the magnetic field suppresses fragmentation in primordial star formation, as well as the present-day star formation.
The models presented here have a small amplitude of the non-axisymmetric density perturbation ($A_{\varphi}=0.01$) at the initial stage.
If larger amplitude was set at the initial stage, the boundary between the fragmentation and non-fragmentation models might change slightly. 
\citet{machida07c} discussed the fragmentation condition for a present-day magnetized cloud, and indicated that the fragmentation condition does not depend qualitatively on the initial amplitude of the non-axisymmetric perturbation.
Moreover, \citet{machida07d} showed that fragmentation condition for a primordial non-magnetized cloud depend slightly on the initial amplitude of the non-axisymmetric perturbation.
We therefore expect that the initial amplitude of the non-axisymmetric perturbation hardly affect the fragmentation condition even in the primordial magnetized collapsing cloud.

\subsection{Jet and Magnetic/Rotational Energy}

\begin{figure}
\begin{center}
\includegraphics[width=150mm]{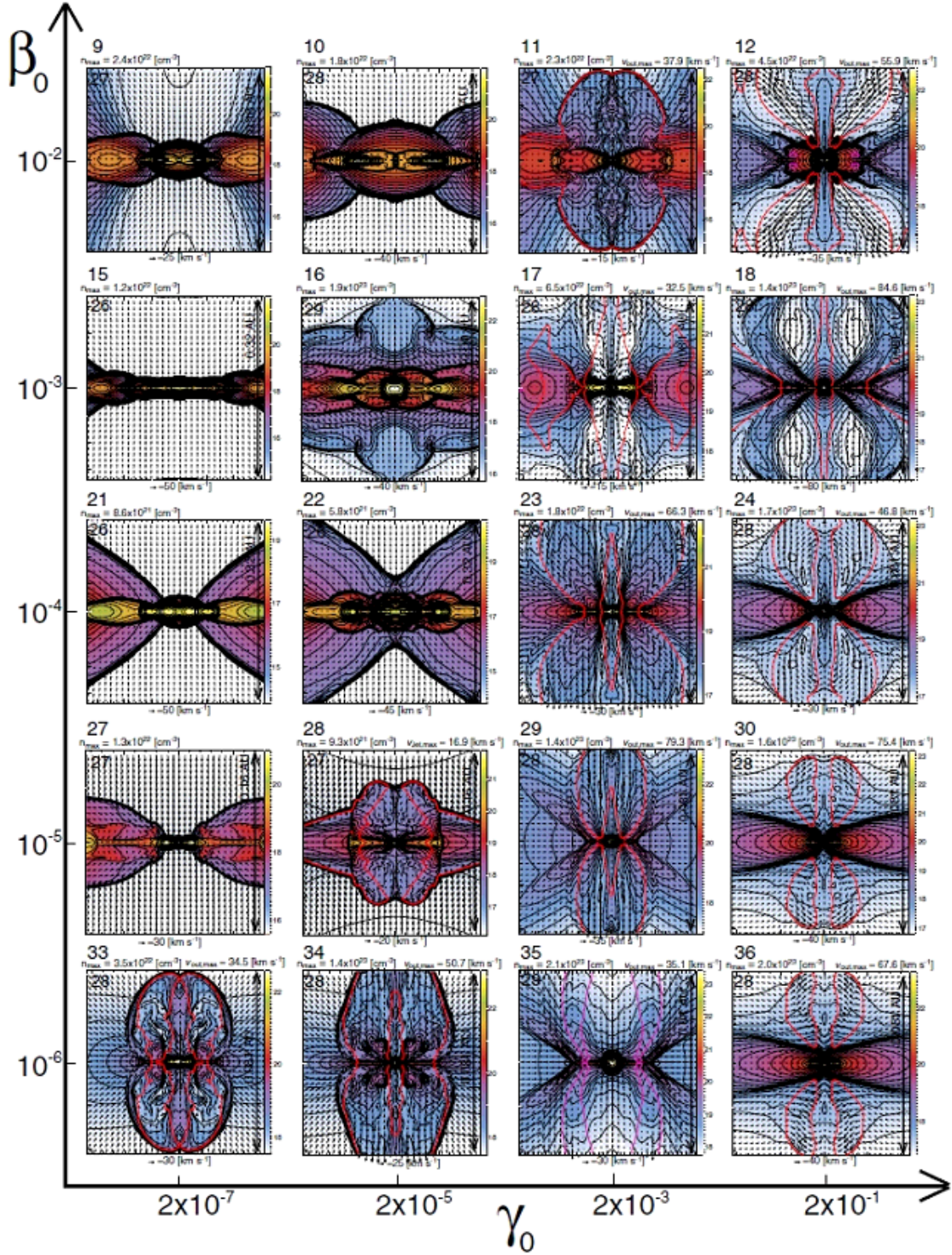}
\caption{
Final states on the $y=0$ plane against parameters $\gamma_0$ and $\beta_0$.
The model numbers are described in the upper left corner outside each panel.
Density ({\it false color} and {\it contours}), velocity vectors ({\it arrows}) are plotted in each panel.
The red line represents the border between the infall and outflow.
The central number density $\nc$, elapsed time $t$, grid level, grid scale and velocity unit are shown in each panel.
}
\label{fig:5}
\end{center}
\end{figure}

Figure~\ref{fig:5} shows final states in the plane of the initial magnetic ($\gamma_0$, $x$-axis) and rotational energies ($\beta_0$, $y$-axis), contrasting the jets at almost the same evolutionary stage of $n_c \simeq 10^{22-23}\,{\rm cm}^{-3}$.
Each panel shows the density and velocity distribution in the $y=0$ plane, and  a thick red contour denotes the boundary between inflow ($v_r <0 $) and outflow ($v_r > 0 $) regions.
The gas flows out of the central region inside the red contour.
The panels without thick red contours indicate the models where a jet does not appear.
As listed in Table~\ref{table}, jets appear in 13 of 36 models.
We call these 13 models `jet models', and other models `non-jet models.'
In all the jet models, a jet appears only after a protostar is formed.
After the protostar formation, the rotation timescale becomes shorter than the collapse timescale because of the accretion of the angular momentum and hardness of the equation of stage at the cloud center, and the strong centrifugal force drives the jet via disk wind mechanism \citep{tomisaka02,banerjee06,machida07b,hennebelle07b}.
The jet disturbs the density distribution near the protostar.
As shown in Figure~\ref{fig:5}, the jet supplies gas above and below the disk, and the density becomes higher there for the jet models, while these region remains less dense for the non-jet models.

The jet models are distributed in the lower right region in Figure~\ref{fig:5}, indicating that a jet is driven in strongly magnetized but slowly rotating cloud.
It should be also noted that the non-jet models coincide with the fragmentation models except for models 31 and 32. 
In other words,  almost all the clouds experience either jet formation or fragmentation.
In the fragmentation models, the angular momentum of the parent cloud is distributed to both the orbital and spin angular momenta of the fragments, and fragmentation therefore reduces the spin angular momentum of the protostar.
This deficiency of the spin angular momentum of the fragments suppresses jet formation.

Firstly, to investigate the relation between jet and magnetic field strength, we focus on models in the fourth row (models 27-30).
In model 27, which has a weakest magnetic field $B_0 = 10^{-8}$\,G  (or $\gamma_0=2\times 10^{-7}$) at the initial stage, no jet appears because of fragmentation.
In model 28, the protostar manages to drive the jet, while the initial cloud has a considerably small magnetic energy ($\beta_0 = 2\times 10^{-5}$).
When the cloud has a weak magnetic field, a jet is considered to be driven by the magnetic pressure gradient force, not by the disk wind mechanism. 
The jet structure in model 28 is similar to the magnetic bubble as reproduced by \citet{tomisaka02}, \citet{banerjee06} and \citet{hennebelle07b}.
\citet{tomisaka02} shows that when the magnetic field around the central object is extremely weak ($\beta_{\rm p} \gg 1$, where $\beta_{\rm p}$ is the plasma beta),  the magnetic field is amplified by the spin of the central object, and the magnetic pressure drives a jet.

A prominent jet is reproduced in model 29, in which the jet is considered to be driven mainly by the disk wind mechanism.
The jet in model 30 is weaker than that in model 29 in spite of the stronger initial field strength; the jet in model 30 has slower average speed and smaller mass ejection rate than those in model 29.
The weakness of the jet in model 30 is attributed to the strong magnetic braking, which slows down the spin of the cloud before the protostar formation in model 30.
In fact, the protostar has an angular velocity of $\Omega_{\rm c}=1.0\times 10^{-5}$\,s$^{-1}$ in model 29 while $\Omega_{\rm c}=1.1\times 10^{-6}$\,s$^{-1}$ in model 30 at the protostar formation epoch.
The protostar therefore rotates $\sim10$ times faster in model 29 than that in model 30.
In addition, the growth rate of the angular velocity in the collapsing cloud is smaller in strongly magnetized cloud (see \S\ref{sec:dis-ang-mag}).
In models 27-30, the jet in model 29 has highest speed and largest jet momentum flux at the end of the calculation.
As a result, moderate strength of the magnetic field is necessary to drive strong jet from the protostar.

Next, to investigate dependence on rotation, we focus on the models in the third column in Figure~\ref{fig:5} (models 11, 17, 23, 29, and 35).
In these models, the most prominent jet appears in model 23, while weak jets appear both in slowly (model 35) and rapidly rotating clouds (model 11).
The weakness of the jet in model 35 is due to deficiency of rotation to drive a strong jet, 
and that in model 11 is due to a low amplification of the magnetic field during the collapse; amplification of the magnetic field is related to the rotation, and spin-up is also related to the magnetic field strength as discussed in \S\ref{sec:dis-ang-mag}.
Therefore, a moderate rotation speed as well as a moderate magnetic field strength is necessary to drive a strong jet.
In summary, a jet appears in models distributed in the lower right region (i.e., models with stronger magnetic field and slower rotation), and models showing strong jets are limited in the range of $10^{-5}\lesssim \beta_0 \lesssim 10^{-3}$ and $10^{-3}\lesssim \gamma_0 \lesssim 10^{-1}$.

Since we calculated the evolution of the jets only for short duration ($\sim$\,a few 10 days) for numerical limitation, we cannot estimate the momentum of a jet, mass ejection speed and final speed of jet in long-term evolution.
In addition, strength of a jet may largely change  in further long-term evolution.
We do not discuss the further evolution of a jet (c.f., final jet speed, mass ejection rate, etc..) , because our purpose in this paper is to investigate the condition for driving jet in primordial collapsing clouds.

\subsection{Configurations of Jets and Magnetic Field Lines}

\begin{figure}
\begin{center}
\includegraphics[width=140mm]{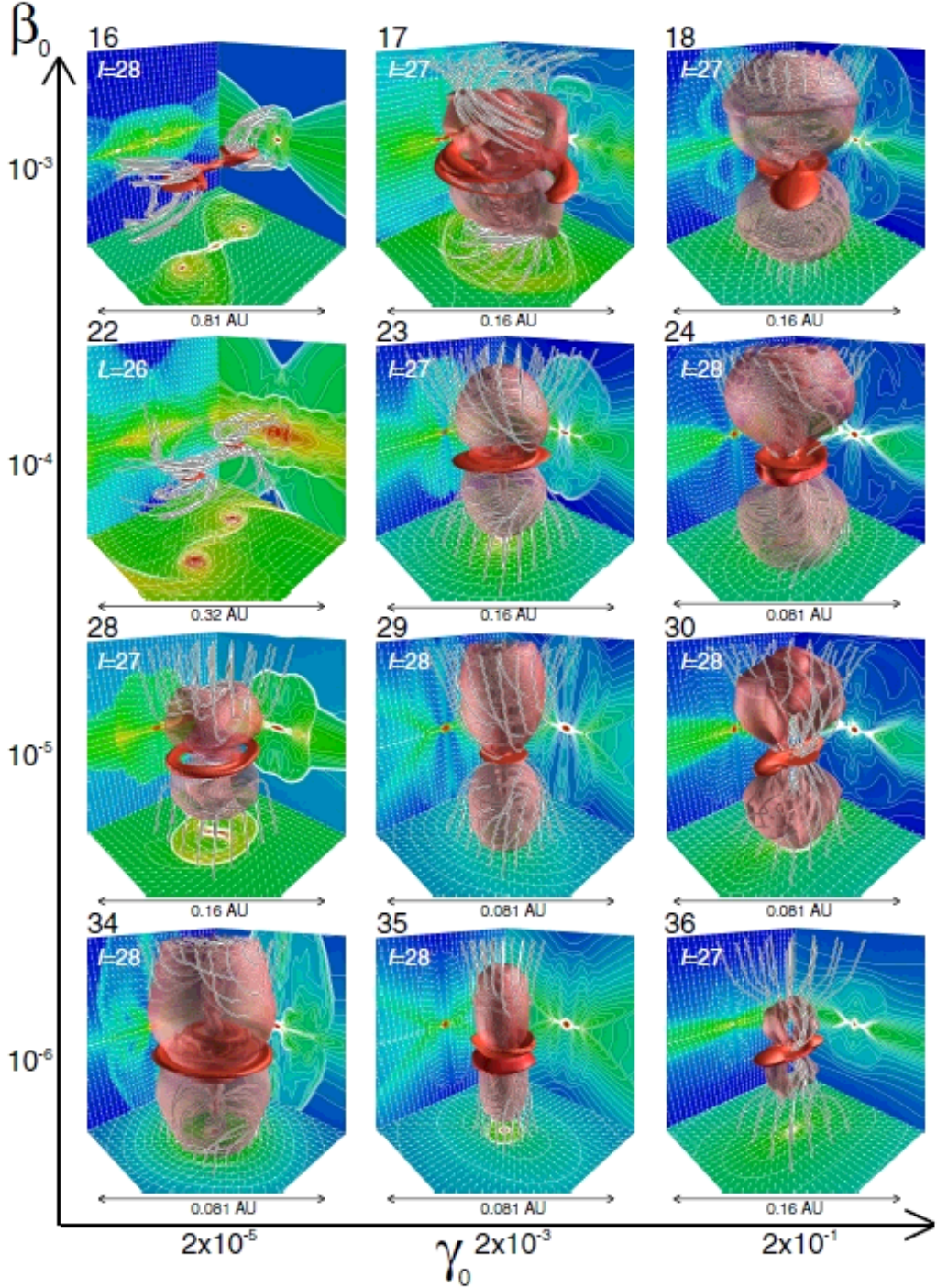}
\caption{
Final states in three dimensions against parameters $\gamma_0$ and $\beta_0$.
The model numbers are described in the upper left corner outside each panel.
The magnetic field lines ({\it black and white streamlines}), high-density regions ({\it red isosurface}), and outflow region ({\it transparent red isosurface}) are plotted in each panel.
The density contours ({\it false color} and {\it contour lines}) and velocity vectors ({\it thin arrows}) are also projected in each wall surface.
The grid level, and grid scale are shown in each panel.
}
\label{fig:6}
\end{center}
\end{figure}

Figure~\ref{fig:6} shows a configuration of magnetic field lines (black-white streamlines) and a structure of a jet (transparent red iso-velocity surface) for each models.

%% jet models (axisymmetric)
In the jet models, the magnetic field lines are twisted inside the jet region.
Almost axisymmetric jets are seen in models 18, 23, 24, 28, 29, 34, and 35.
These jets have the hourglass-like configurations of the magnetic field lines, where the poloidal component is more dominant than the toroidal component as shown in Figure~\ref{fig:1}{\it f} and \ref{fig:2}{\it c}.
These configurations of the magnetic field lines can easily drive a jet by the disk wind mechanism \citep[e.g.,][]{blandford82}.
%% jet models (non-axisymmetirc)
Non-axisymmetric jets are seen in models 17, 30, and 36.
They are caused by the non-axisymmetric density distribution at the protostar formation epoch.
Such a non-axisymmetric density distribution is also reflected by non-axisymmetric circumstellar disks, represented by a red iso-density surface in Figure~\ref{fig:6}.
For example, model 30 exhibits the non-axisymmetric jet (transparent red iso-velocity surface) driven from the bar-like density distribution (red iso-density velocity surface) at the root of the jet.

%% fragmentation models
In the fragmentation models, the configurations of the magnetic field lines are disturbed, and the toroidal field tends to be more dominant than the poloidal filed as shown in models 16 and 22 in Figure~\ref{fig:6} and also in Figure~\ref{fig:3}{\it f}.
These configurations of the magnetic field lines indicate weakness of the field strength, which is insufficient to drive a jet.
The weak poloidal field is attributed to the oblique shocks on the surface of the disk envelope, and the disturbance of the field lines are attributed to the orbital motion of the fragments and the spin of the fragments.
The spin of the fragment winds the magnetic fields around its rotation axis, and the field strength is amplified. 
In the further stages, the magnetic pressure may drive a jet. 
%%We should also comment on an advantage of the weak poloidal field and the strong toroidal field in fragmentation, where the magnetic tension and magnetic pressure suppresses merger of fragments as indicated in \citet{price07}.

\section{DISCUSSION}
\subsection{Magnetic Flux-Spin Relation}
\subsubsection{Evolution Track for Magnetic Field and Angular Velocity}
\label{sec:dis-ang-mag}
Figure~\ref{fig:7} shows the evolution track of the central magnetic field strength ($x$-axis) and central angular velocity ($y$-axis)  from the initial state for some models.
The $x$-axis indicates $ B_c/(8\pi P_c)^{1/2}$, which corresponds to the pressure ratio of $\left(P_{\rm mag}/P_c\right)^{1/2} $ at the center, where $P_{\rm mag}$ and $P_c$ denote the central magnetic pressure and the central thermal pressure.
This variable also corresponds to the inverse square root of the plasma beta, $\beta_{\rm p}^{-1/2}$.
Hereafter this variable is called the normalized magnetic field.
The $y$-axis indicates $\Omega_{\rm c}/(4\pi G \rho_c)^{1/2}$, which corresponds to a ratio of the time scales, $t_{\rm ff}/t_{\rm rot}$, at the center, where $t_{\rm ff}$ and $t_{\rm rot}$ denote the free-fall and rotation time scales.
This variable also coincide with the energy ratio of $\left( E_{\rm rot} / | E_{\rm grav}| \right)^{1/2}$ at the center if assuming a uniform sphere.
Hereafter this variable is called the normalized angular velocity.
The asterisks at the ends of the loci denote the initial values of normalized magnetic fields and normalize angular velocities.
The diamonds on the loci denote the stages of $\nc = 10^5$, $10^7$, $\cdots$, $10^{19}\cm$, and these values are labeled for the loci of models 2, 9, 21, 23, 33, and 35.
The arrows on the loci indicate the directions of the cloud evolutions.

The normalized magnetic fields and angular velocities increase as the central density increases except for stages of very high density in all the models.  
For comparison, a relationship of 
\begin{equation}
\dfrac{\Omega_{\rm c}}{(4\pi G \rhoc)^{1/2}} \propto \dfrac{B_{\rm c}}{(8 \pi P_{\rm c})^{1/2}}.
\label{eq:solid}
\end{equation}
is plotted with the solid line at the lower right corner of Figure~\ref{fig:7}, and all the model evolves in the direction almost parallel to this line. 
This implies that both of the normalized magnetic fields and the normalized angular velocities are amplified during the early phase of the collapse at the almost same growth rate.
As indicated by \citet{machida05a,machida06a,machida07a}, the central magnetic field and angular velocity increase as,
\begin{equation}
B_c, \ \Omega_c \propto \rho_c^{2/3}
\label{eq:mag-omega}
\end{equation}
during the nearly spherical collapse phase that corresponds to Figure~\ref{fig:1}{\it a}--\ref{fig:1}{\it d}.
%%, which is a early phase of the cloud collapse, for example, as shown in 
The relationship of the angular velocity in equation~(\ref{eq:solid}) is rewritten as, 
\begin{equation}
\dfrac{\Omega_{\rm c}}{(4\pi G \rhoc)^{1/2}} \propto \rho_c^{1/6},
\label{eq:omega}
\end{equation}
and the relationship of the magnetic field is rewritten as,
\begin{equation}
\dfrac{B_{\rm c}}{(8 \pi P_{\rm c})^{1/2}} \propto \rho_c^{0.12}.
\label{eq:bz}
\end{equation}
The power index in the right hand side of (\ref{eq:bz}) is slightly lower than that in equation (\ref{eq:omega}) although they coincide in the isothermal case.
The difference between the power indexes is attributed to the equation of state in the primordial gas, $P = \rho^\gamma$. 
The polytropic index $\gamma$ is assumed as a function of density, and it is approximated to $\gamma \approx 1.1$ for $10^4\cm \lesssim n \lesssim 10^{17}\cm$ (see Fig.1 of \citealt{machida07a}).
Adopting the relation of  $P_c \propto \rho_c^{1.1}$, the equation (\ref{eq:bz}) is obtained. 
The slightly larger power index in the right hand side of equation (\ref{eq:bz}) is reflected in the slope of loci slightly steeper than that of equation (\ref{eq:solid}).
The relationships of equations (\ref{eq:omega}) and (\ref{eq:bz}) are examined in detail in \S\ref{sec:dis-spherical-collapse}.

The slope of loci also indicates that the magnetic braking is ineffective during the spherical collapse phase.  
If magnetic braking is effective (i.e., it slows down the spin of a cloud), increase in the angular velocity was less than that indicated by equation (\ref{eq:omega}), and the slope of the loci would be shallower.
We confirmed that the magnetic braking become prominent only after the protostar formation. 
In other words, the magnetic field is not amplified enough to affects the evolution of the cloud during the spherical collapse of the early phase.
In the later phase, the cloud center exhibits transition from nearly spherical collapse to the disk-like collapse because of rapid rotation and/or strong magnetic field, which are amplified in the nearly spherical collapse phase, and the evolutionary track stagnated in the hatched band of Figure~\ref{fig:7}.
The disk-like collapse phase is discussed in \S\ref{sec:dis-convergence}.

\subsubsection{Magnetic Field and Angular Velocity in the Spherical Collapse Phase}
\label{sec:dis-spherical-collapse}
To verify the relation of equations~(\ref{eq:omega}) and (\ref{eq:bz}), the evolutions of the central angular velocities and magnetic fields are plotted against the central density in Figures~\ref{fig:8} and \ref{fig:9}, respectively.

Figure~\ref{fig:8} shows the evolution of the normalized angular velocity in models 3, 9, 15, 21, 27, and 33, which are located in the third column of Figure~\ref{fig:4} and have the common weak magnetic field $B_0 = 10^{-7}$\,G and different angular velocities $\Omega_0=2.3\times 10^{-17}-7.2\times 10^{-15}$\,s$^{-1}$ at the initial stage (see Table~\ref{table}).
In Figure~\ref{fig:8}, the normalized angular velocities in all models except for model 3 evolve with $\propto \rho^{1/6}$ from the initial stage, indicating the power law amplification of $\Omega_c \propto \rho_c^{2/3}$ and spherically collapse of the clouds.
On the other hand, the cloud in model 3 cannot collapse spherically because the cloud has a large rotational energy at the initial stage.

Figure~\ref{fig:9} shows the evolution of the magnetic field strength normalized by the square root of the density (Fig.~\ref{fig:9} upper panel) and the normalized magnetic field (Fig.~\ref{fig:9} lower panel) in models 32-36.
Models 32-36 are located in the sixth row in Figure~\ref{fig:4}, and have the common small angular velocity $\Omega_0 = 2.3\times 10^{-17}$\,s$^{-1}$ and different strengths of the magnetic field $B_0= 10^{-9}-10^{-6}$\,G at the initial stage  (see Table~\ref{table}).
The upper panel indicates the power law amplification of $B_c \propto \rho_c^{2/3}$ (equation~[\ref{eq:mag-omega}]) except in the high-density phase and therefore the spherical collapse of the clouds for all models.
Figure~\ref{fig:9} lower panel also shows the evolution of the magnetic field in terms of the $\beta_{\rm p} ^{-1/2} (= B_c/\sqrt{8 \pi P_c})$.
All the curves indicate the power law amplification of $\beta^{-1}\propto \rho^{0.12}$, corresponding to equation~(\ref{eq:bz}).
The curves in the lower panel undulate more than those in the upper panel because of the undulation in the pressure included in the normalization.
However, the curves in the lower panel converges well to $\beta_{\rm p}^{-1/2} \simeq 0.2$ at the high density phase compared with those in the upper panel.
As a result, when the cloud collapses spherically (i.e., early evolution phase), the evolution of the angular velocity and magnetic field are well expressed by equation~(\ref{eq:mag-omega}),  and the normalized variables are also expressed by equations~(\ref{eq:omega}) and (\ref{eq:bz}).

\subsubsection{Convergence of the Magnetic Field and Angular Velocity}
\label{sec:dis-convergence}
\begin{figure}
\begin{center}
\includegraphics[width=160mm]{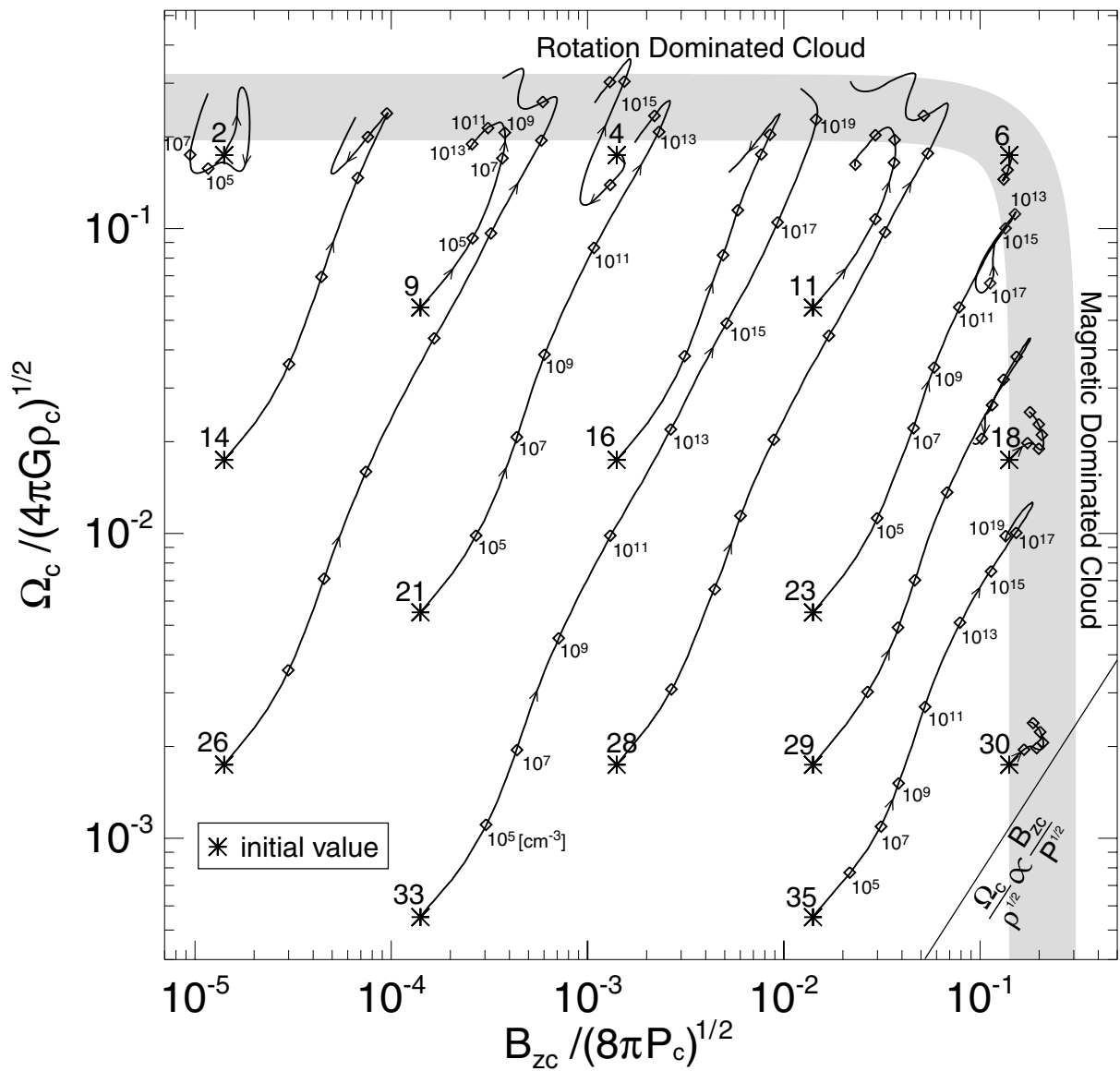}
\caption{
Magnetic flux--Spin relation.
The evolution of magnetic flux density and angular velocity at the cloud center are plotted.
The $x$-axis indicates the square root of the magnetic pressure [$B_c\,/ (8\pi)^{1/2}$] divided  by the square root of the thermal pressure ($ P_{\rm c}$). 
The $y$-axis represents the angular speed ($\Omega_c$) divided by the free-fall rate [$(4 \pi G \rho_c)^{1/2}$].
The symbols represent the magnetic field and the angular velocity at the initial state ($*$) and at each central density ($\diamond$: $\nc=10^5, 10^8, 10^{11} \cdot \cdot  \cm$).
Each line denotes the evolution path from the initial state ($n_{\rm c,0} = 1.86\times 10^3 \cm$).
The arrows mean the direction of evolution.
The thick grey band denotes the magnetic flux--spin relation $\Omega_c^2/[(0.25)^2 \times 4\pi G \rho_c] + B_{c}^2/[(0.2)^2 \times 8\pi P_{\rm c}] =1$ [see eq.~(\ref{eq:bw})].
}
\label{fig:7}
\end{center}
\end{figure}

\begin{figure}
\begin{center}
\includegraphics[width=160mm]{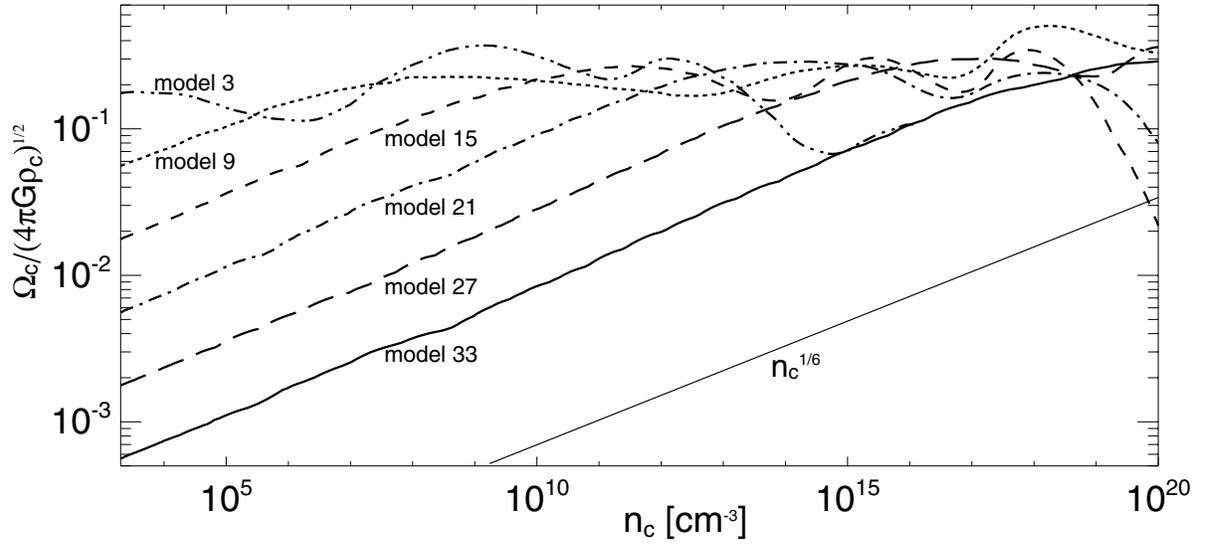}
\caption{
Angular velocity ($\Omega_{\rm c}$) normalized by the freefall timescale $(4\pi G \rho_{\rm c} )$ against the central number density $n_{\rm c}$ for models 3, 9, 15, 21, 27, and 33.
The relation $\Omega_{\rm c}/(4\pi G \rho_{\rm c})^{1/2} \propto \nc^{1/6}$ is also plotted.
}
\label{fig:8}
\end{center}
\end{figure}

\begin{figure}
\begin{center}
\includegraphics[width=160mm]{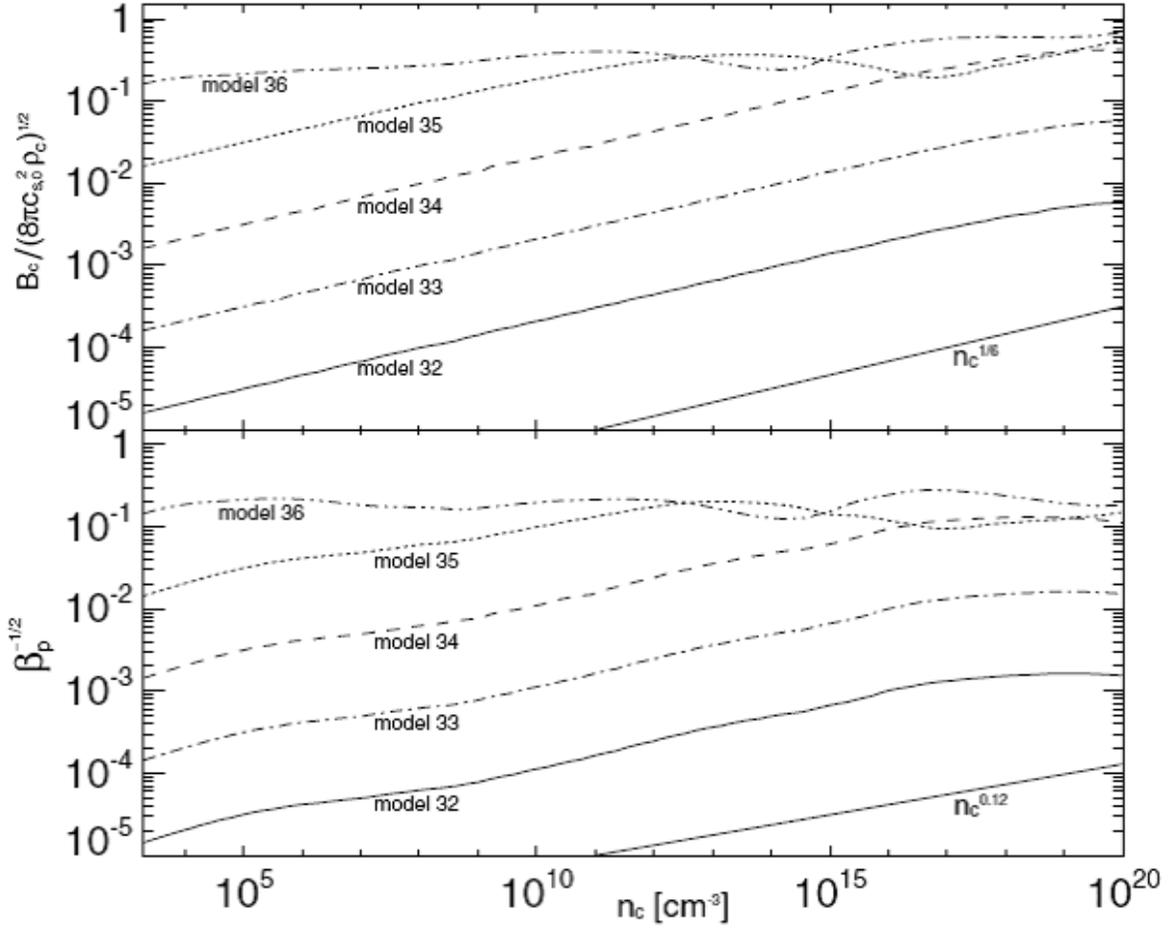}
\caption{
The magnetic field strength normalized by the square root of the density [$B_{\rm z}/(8\pi c_{\rm s,0}^2 P_{\rm c})^{1/2}$] (upper panel), where $c_{s,0}$ is the initial sound speed, and inverse root of the plasma beta $\beta_{\rm p}^{-1/2}$ (lower panel) are plotted against the central number density $n_{\rm c}$ for models 31--35.
The relations $B_c/(4\pi c_{\rm s}^2\rhoc)^{1/2}\propto n_{\rm c}^{1/6}$ (upper panel) and $\beta_{\rm p}^{-1/2} \propto n_{\rm c}^{0.12}$ (lower panel) are also shown.
}
\label{fig:9}
\end{center}
\end{figure}

As the cloud collapses, the normalized angular velocities converge to $\Omega_c/(4\pi G \rhoc)^{1/2} \simeq 0.25$ in Figure~\ref{fig:8}, and the normalized magnetic fields converge to $B_c/(8\pi P_c)^{1/2}\simeq 0.2$ with initially strong magnetic fields (models 34, 35, and 36) in Figure~\ref{fig:9} lower panel.
Figures~\ref{fig:8} and \ref{fig:9} present only limited models.
As shown in Figure~\ref{fig:7} where more models are plotted, the normalized magnetic field and the normalized angular velocity converge to the so-called flux-spin relation,
\begin{equation}
   \frac{\Omega_{c}^2}{(0.25)^2 \, 4 \pi G \rho_c} + \frac{B_{\rm c}^2}{(0.2)^2 \, 8 \pi P_{\rm c}} = 1.
\label{eq:bw}
\end{equation}
This relation is represented by the gray band in Figure~\ref{fig:7}.
If an initial stage is located in lower and left-side of the gray band in Figure~\ref{fig:7}, a points indicated by the two normalized variables moves in the upper right direction according to equations (\ref{eq:omega}) and (\ref{eq:bz}) in the nearly spherical collapse phase. 
When the point reaches the gray band of equation (\ref{eq:bw}), the cloud changes the geometry of collapse from a sphere to a disk. 
After the transition, the point remains around the gray band with small oscillation.

The numerical factors in equation (\ref{eq:bw}) are determined empirically.
The horizontal gray band corresponds to the convergence value in Figure~\ref{fig:8} [$\Omega_{\rm c}/({4\pi G \rhoc})^{1/2} \simeq 0.25$], and the vertical gray band corresponds to the convergence value in Figure~\ref{fig:8} lower panel [$B_{\rm c}/(8\pi P_{\rm c})^{1/2}\simeq 0.2$].
The convergence value of the angular velocity was found by \citet{matsu97}, and that of magnetic field was found by \citet{nakamura99}, for isothermal clouds modeling the present-day star formation.
While they considered either rotation or magnetic field, \citet{machida05a, machida06a} considered both the rotation and magnetic field, and obtained the convergence values for rotation and magnetic field (flux-spin relation) for the present-day star formation.
Equation~(\ref{eq:bw}) is an extension of flux-spin relation of \citet{machida05a, machida06a} to the primordial star formation.
Figure~\ref{fig:7} indicates that the magnetic field strengths and angular velocities converge to the flux-spin relation represented by equation~(\ref{eq:bw}) even in the primordial magnetized clouds.

The length of the locus before the convergence to the spin-flux relation depends on the initial magnetic field strength and the initial angular velocity.
Figures~\ref{fig:8} and \ref{fig:9} show that models having stronger magnetic field and/or more rapid rotation at the initial state converge to the spin-flux relation in the earlier evolutionary phase (or lower density phase).
In other words, a model initially located near the gray band  converges earlier.
For example, in Figure~\ref{fig:7}, the evolutionary track of model 9 reaches gray band at $\nc \simeq 10^{9}\cm$, while that of model 21 reaches the gray band at $\nc \simeq 10^{13}\cm$.
This is because the initial normalized angular velocity in model 9 is stronger than that in model 21 (models 9 and 21 have the same strengths of the magnetic field at the initial).
In order to confirm the relation of equations~(\ref{eq:omega}) and (\ref{eq:bz}) in Figure~\ref{fig:7}, we focus on the evolutionary track for model 33, which has $\Omega_c/(4\pi G\rho_c)^{1/2}=5.5\times 10^{-4}$  and $B_c/(8\pi c_{s}^2\rho_c) =1.4\times 10^{-4}$ at the initial stage.
The evolutionary track in model 33 reaches gray band at $n\simeq 10^{19}\cm$, that is $10^{16}$ times higher  than the initial density. 
According to equations~(\ref{eq:omega}) and (\ref{eq:bz}), the normalized angular velocity and magnetic field should be amplified by factors of 464 and 83.2  respectively.
On the other hand, the model 33 exhibits $\Omega_c/(4\pi G\rho_c)^{1/2}\simeq 0.2$ and $B_c/(8\pi c_{s}^2\rho_c) =1\times 10^{-2}$ when the evolutionary track reaches gray band, and the amplification factors are therefore 364 and  71.4, which are consistent with that derived from equations~(\ref{eq:omega}) and (\ref{eq:bz}).

When an evolutionary track converges to the flux-spin relations, the central cloud deforms from a sphere to a disk because the central angular velocity and/or magnetic field have been amplified during the nearly spherical collapse phase.
For example, model 23, shown in Figure~\ref{fig:1}, produces a thin disk at $\nc \simeq 10^{13}\cm$, at which the evolutionary track converges to the gray band in Figure~\ref{fig:7}.
After evolutionary track converges to the flux-spin relation, it oscillates around  the flux-spin relation as shown in Figures~\ref{fig:7}.  
We also noticed that the strong accretion shocks are generated on the surfaces of the disk when the model converges to the flux-spin relation. 
The shock generation is also related to the oscillation of the evolutionary track after the convergence.
When the evolutionary track oscillates around the flux-spin relation, a new shock is generated inside the collapsing disk intermittently.
The intermittent shock generation results in a nested structure of the shocks.
Such nested shocks are first reproduced in the simulation of collapse of a rotating isothermal disk by \citet{norman80}.
The convergence and oscillation is also seen in Figures~\ref{fig:8} and \ref{fig:9}.

The convergence indicates a power law amplification of angular velocity and magnetic field during the disk-like collapse phase.
In the case of homologous collapse of an isothermal disk with a thin disk approximation, 
this power law amplification is expressed as $\Omega_c, B_c \propto \rhoc^{1/2}$ \citep{machida05a}.
For the polytropic gas, the growth rate is modified as $\Omega_c, B_c \propto \rho_c^{\gamma/2}$. 
The normalized variables are also modified as $\Omega_c/(4\pi G \rho_c)^{1/2} \propto \rho_c^{(\gamma -1)/2}$ and $B_c/(8\pi P_c)^{1/2} = {\rm constant}$ \citep[see also][]{machida07d}.
In our case of $\gamma \approx 1.1$, the normalized angular velocity would increase slightly in proportion to $\rho_c^{0.05}$, and such a small increase is hardly seen in Figures~\ref{fig:7} and \ref{fig:8} because of the significant amplitude of the oscillation.

\subsection{Fragmentation/Jet Condition}
\label{sec:dis-frag-jet-cond}
All the models examined in this paper exhibit convergence to the flux-spin relation of equation~(\ref{eq:bw}) as shown in Figure~\ref{fig:7}.
A convergence point within the flux-spin relation depends on the initial angular velocity and the initial magnetic field, and the convergence point characterizes the fate of a cloud. 
A model with initially fast rotation and weak magnetic field starts its evolutionary track at a point in the upper left region, and the evolutionary track converges to the horizontal gray band.
Such a model is called ``the rotation-dominated model.''
On the other hand, a model with initial slow rotation and strong magnetic field starts its evolutionary track at a point in the lower right region of Figure~\ref{fig:7}, and the evolutionary track converges to the vertical gray band.
Such a model is called ``magnetically-dominated model.''
We classified all the models into the rotation ($\Omega$) and magnetic ($B$) dominated models and summarized it in the last column of Table~\ref{table}.

Only the rotation-dominated models exhibit fragmentation because the rotation is amplified enough to promote fragmentation and magnetic braking is insufficient to prevent fragmentation.
\citet{machida07d} examined fragmentation of non-magnetized primordial clouds, and showed that fragmentation occurs in the case where the normalized angular velocity at the center  converges to 
\begin{equation}
\dfrac{\Omega_c}{\sqrt{4\pi G \rhoc}} \simeq 0.25
\label{eq:fcri}
\end{equation}
before the protostar formation.
A magnetically-dominated model never converges to equation~(\ref{eq:fcri}) because the evolutionary track never reaches the horizontal band in Figure~\ref{fig:7}.
A rotation-dominated model is satisfied with equation~(\ref{eq:fcri}) when the initial normalized angular velocity is 
\begin{equation}
\dfrac{\Omega_0}{\sqrt{4\pi G \rho_0}} \gtrsim 0.25 \left(\frac{\rho_0}{\rho_{\rm cr}}\right)^{1/6},
\label{eq:fcri2}
\end{equation}
where $\rho_{\rm cr} \simeq 10^{17} \cm$ denotes an upper-bound of density where the equation state is approximated by the polytrope with $\gamma \approx 1.1$, and $\rho_0 = 1.86 \times 10^3 \cm$ denote the initial central density.
We used the relationship of $\Omega_c \propto \rho_c^{2/3}$ when equation~(\ref{eq:fcri2}) is derived.
Examining the initial conditions of the rotation-dominated models, all the rotation-dominated models except for models 31, 32, and 33 satisfies equation~(\ref{eq:fcri2}), and these models exhibit fragmentation, consistently with the fragmentation condition.
The fragmentation condition of equation~(\ref{eq:fcri}) is therefore valid even for magnetized primordial clouds.

The fragmentation condition can be rewritten in terms of the rotational energy.
Assuming that the center of the cloud undergoes homologous collapse with rigid rotation, the parameter $\beta (=E_{\rm rot}/|E_{\rm grav}|)$ is expressed as $\beta=\Omega^2/(4\pi G\rho)$.
Equation~(\ref{eq:fcri}) therefore indicates that fragmentation occurs when the rotational energy reaches $\sim6$\% of the gravitational energy owing to spinning up of the cloud center.

All the magnetically-dominated models presented here show the jet formation.
The condition for jet formation coincides with the flux-spin relation of the magnetic-dominated models, 
\begin{equation}
\dfrac{B_c}{\sqrt{8\pi P_c}} \simeq 0.2.
\label{eq:out-cond}
\end{equation}
When the central magnetic field is amplified up to the level of equation~(\ref{eq:out-cond}) before the protostar formation, a jet appears just after a protostar is formed. 
This condition is expressed as, 
\begin{equation}
\dfrac{B_0}{\sqrt{8\pi P_0}} \gtrsim 0.2 \left(\frac{\rho_0}{\rho_{\rm cr}}\right)^{0.12},
\label{eq:out-cond2}
\end{equation}
using the power law amplification of the magnetic field during the spherical collapse (eq. [\ref{eq:bz}]). 
Examining the initial conditions of the magnetically-dominated models, they are satisfied with equation~(\ref{eq:out-cond2}) except for model 34, although the evolutionary track of model 34 converges to the point satisfying equation~(\ref{eq:out-cond}). 
This exceptional feature of model 34 is attributed to the slightly steeper slope in the lower panel of Figure~\ref{fig:8} than that of the reference line, $\beta_{\rm p}^{-1/2} \propto \nc^{0.12}$.
Equation~(\ref{eq:out-cond}) can be rewritten as $\beta_{\rm p} \simeq 25$ since $B_{\rm c}/(8\pi P)^{1/2} = \beta_{\rm p}^{-1/2}$.
This indicate that a jet is driven by the protostar when the magnetic field is amplified  and the plasma beta reaches $\beta\simeq 25$ before the protostar formation.
The equation~(\ref{eq:out-cond}) cannot be realized in rapidly rotating clouds, because the evolutionary track converges to horizontal gray band in Figure~\ref{fig:7} and never reach the vertical band.

Although the magnetically and rotation dominated models are separated clearly, some models are located on the border between the categories present mixed features. 
Model 11 and 17 are located near the border between the rotational and magnetic dominated models, and exhibit both fragmentation and jet.
In these models, fragments merge to form a single core after fragmentation, while other fragmentation model does not show merger until the end of the calculation.
After the merger, the jet begins to be driven from the merged core.

In the fragmentation models, we do not observe a jet driven from either fragment.
This indicates that fragmentation occurs in rotation dominated models where the magnetic field is too weak to drive the jet. 
However, it may be possible that the magnetic field amplified by the spin of the protostars drive the jet in the further stage.
We conclude that the condition of equation~(\ref{eq:out-cond}) is valid for the jet which is formed within a few times 10 days after the protostar formation.

\subsection{Magnetic Fields and Rotation Periods of Proto-Population III Stars}
In our calculation, both the normalized angular velocity and normalized magnetic field are bounded by values expressed by equations~(\ref{eq:fcri}) and (\ref{eq:out-cond}) (i.e., gray band in Fig.~\ref{fig:7}).
Therefore, given a central density, the maximum angular velocity and magnetic field strength are expressed as,
\begin{equation}
\Omega_{\rm max} \le 0.25 \sqrt{4\pi G \rhoc},
\label{eq:omega-max}
\end{equation}
and 
\begin{equation}
B_{\rm max}  \le 0.2 \sqrt{8 \pi P_c}.
\label{eq:bps}
\end{equation}
When the protostar is formed at $n_c=10^{21}\cm$, which is $10^4$ times larger than $n_{\rm cr}$, the protostar has a maximum angular velocity of  
\begin{equation}
\Omega_{\rm ps} = 1.1\times 10^{-5} \  {\rm s}^{-1}, 
\label{eq:wps}
\end{equation}
which corresponds to the rotation period of $P>6.6$\,days.
The maximum magnetic field strength is given by,
\begin{equation}
B_{\rm ps}  = 2.6 \times 10^4 \ {\rm G}.
\label{eq:bps}
\end{equation}
at the protostar formation epoch ($n=10^{21}\cm$), where the thermal pressure adopted in our calculation is used.
We measure the magnetic field strengths ($B_{\rm ps}$) and angular velocities ($\Omega_{\rm ps}$) at the protostar formation epoch for every models, and the measured values are listed in 7th and 8th column of Table~\ref{table}.
For the models with parenthesis, we cannot follow the cloud evolution by the protostar formation epoch because fragmentation occurs in relatively early phases.
For these models, the magnetic field and angular velocity just before fragmentation are listed.
As listed in Table~\ref{table}, the protostars at their formation epoch have the angular velocities in the range of  $3.3\times 10^{-7}\,{\rm s}^{-1} \lesssim \Omega_{\rm ps} \lesssim 1.1\times 10^{-5}\,{\rm s}^{-1}$, which corresponds to the rotation periods of $6.6\,{\rm days} \lesssim  P \lesssim 220\,{\rm days}$.
These angular velocities are well bounded by the value of equation~(\ref{eq:wps}).
The magnetic fields are ranged in  $0.01\,{\rm G} < B_{\rm ps} < 3.1\times 10^4\,{\rm G}$,  which are also bounded by the value of equation~(\ref{eq:bps}).

The observations indicate that the present-day protostar have magnetic fields of $\sim1$\,kG at the maximum \citep{johns99a,johns99b, johns01,bouvier06}.
According to our simulations, Population III protostars can possess 10 times stronger magnetic field than protostars at present day.
The smallness of the magnetic field in the present day protostars is attributed to the magnetic dissipation during the protostellar collapse.
\citet{machida07a} studied collapse of magnetized clouds for the present-day star formation, and showed that the magnetic fields of the protostar is $\sim$\,kG at the maximum because the magnetic field is largely dissipated by the Ohmic dissipation in the late phase of collapse, $10^{11}\lesssim n \lesssim 10^{15}\cm$ (see also, \citealt{nakano02}).
On the other hand, the dissipation of the magnetic field is not supposed to be effective in primordial collapsing clouds as shown in \citet{maki04,maki07}.
Population III stars therefore possess stronger magnetic field than that in present day if they are formed via the process of the magnetically-dominated models. 

The magnetic fields and angular velocities of the protostar listed in Table~\ref{table} are the values at the moment of the protostar formation, in which the mass of protostars is only $\sim 10^{-3}-10^{-2}\msun$.
Since stars acquire a large fraction of mass in a subsequent accretion phase, the magnetic field strength and angular velocity may be changed in further evolution.
The angular momentum is transferred by magnetic interaction between the protostar and circumstellar disk.
The magnetic field can be amplified by convection inside the protostar.
However, since the purpose of this paper is to investigate the magnetic effect in collapsing primordial clouds, we do not discuss subsequent evolution of the magnetic field and angular velocity.
To determine the magnetic field and angular velocity of Population III stars, further long-term calculations is necessary including a model of stellar evolution.

\section{SUMMARY}
\label{sec:summary}
\begin{figure}
\begin{center}
\includegraphics[width=160mm]{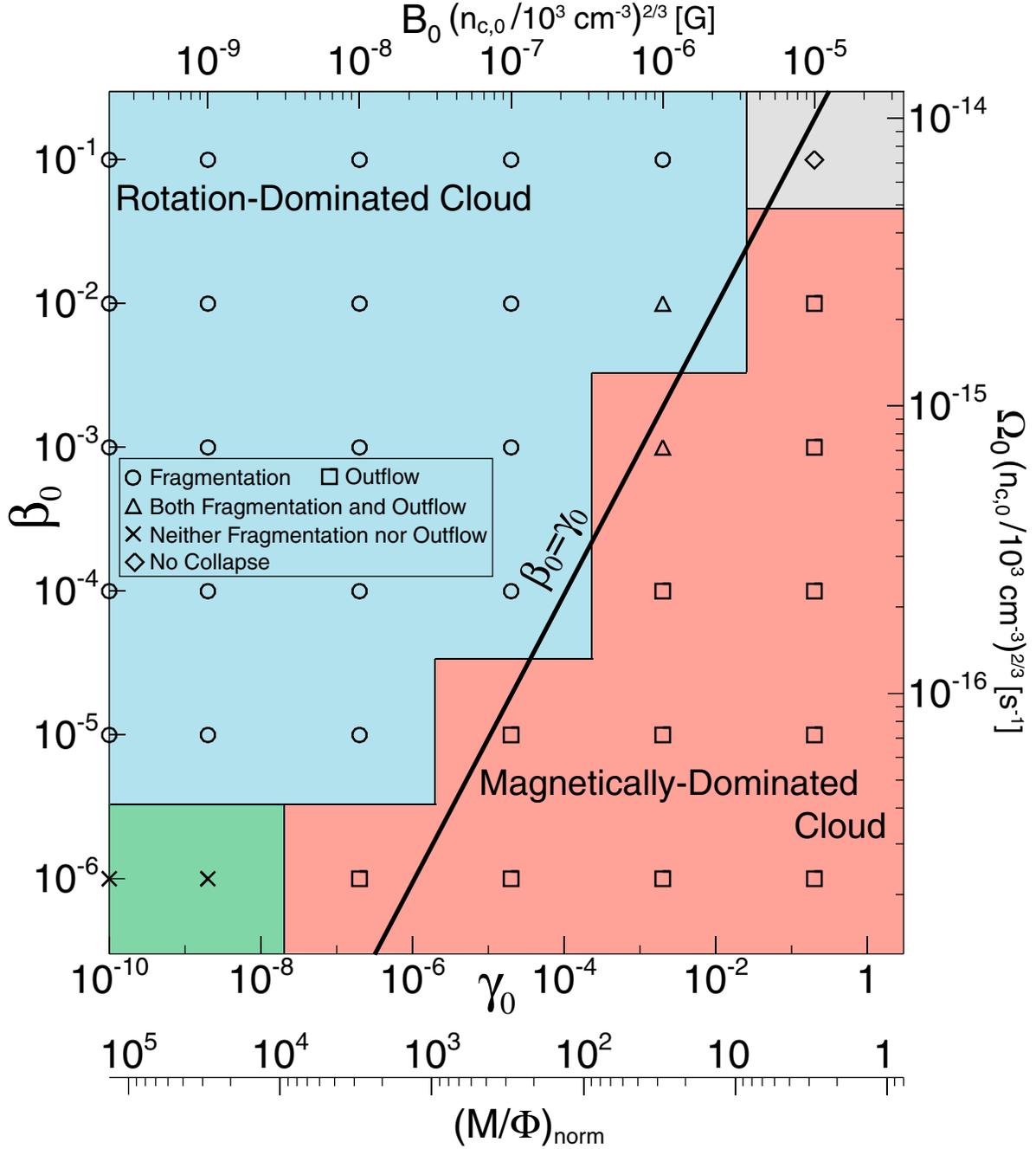}
\caption{
Fragmentation and jet conditions for parameters $\gamma_0$ and $\beta_0$.
Symbols mean the models showing fragmentation ($\circ$), jet ($\square$), both fragmentation and jet ($\triangle$), neither fragmentation nor jet ($\times$), and no collapse ($\diamond$).
The upper and right axes indicate the initial magnetic field strength $B_0$, and  angular velocity $\Omega_0$ depending on the initial density $(n_{c,0}/10^3\cm)^{2/3}$, respectively.
The thick line represent $\beta_0=\gamma_0$.
Bottom axis means the mass to magnetic flux ratio (M/$\Phi$)$_{\rm norm}$ normalized by its critical value.
}
\label{fig:10}
\end{center}
\end{figure}

In this paper, we calculated cloud evolution from the stage of $n_c = 10^3\cm$ until the protostar is formed ($\simeq 10^{22}\cm$) for 36 models, parameterizing the initial magnetic field strength and the initial rotation, to investigate effects of magnetic fields in collapsing primordial clouds.
In figure~\ref{fig:10}, the fates of all the clouds are plotted in the plane of the parameters $\gamma_0$ and $\beta_0$, where the circles, squares, triangles, and crosses mean the models showing fragmentation, jet, both fragmentation and jet, and neither fragmentation nor jet, respectively.
In model located in the upper right corner (model with diamond), the cloud oscillates around the initial state without collapse, because strong magnetic field and rapid rotation suppress the collapse of the cloud.
The solid line in Figure~\ref{fig:10} represents $\beta_0 = \gamma_0$, indicating that the magnetic energy is more dominant than the rotational energy in the model distributed above the line, and vice versa.
All the models showing fragmentation are distributed above the solid line, while almost all the models showing jet are distributed below the solid line.
The solid line clearly separates the models;
fragmentation occurs but no jet appears when $\beta_0 > \gamma_0$, and jet appears after without fragmentation when $\beta_0 < \gamma_0$.
In addition, the solid lines almost coincide with the boundary between the rotation-dominated model and the magnetically dominated models (see Table~\ref{table} for which category each model falls into).
As a result, in the collapsing primordial cloud, the cloud evolution is mainly controlled by the centrifugal force than the Lorentz force when $\beta_0 > \gamma_0$, while the Lorenz force is more dominant than the centrifugal force when $\gamma_0  > \beta_0$.

The upper and right axes of Figure~\ref{fig:10} mean the magnetic field strength and angular velocity of the initial cloud, respectively.
They can be scalable at any initial density as $(\nc /10^3\cm)^{2/3}$, assuming the spherical collapse, which is approximated well in the early phase. 
A jet is driven when the initial cloud has magnetic field of $B_0\gtrsim 10^{-9}(\nc/10^3\cm)^{2/3}$\,G  if the cloud rotates slowly as $\Omega \lesssim 4\times 10^{-17}(\nc/10^3\cm)^{2/3}$\,s$^{-1}$.
This condition corresponds to that in \citet{machida06d} and is also consistent with the condition of equation~(\ref{eq:out-cond2}).

A strong jet is expected in primordial star formation.
The power of a jet, e.g., a mass ejection rate, is considered to be controlled by the accretion rate as indicated in present-day star formation; the mass ejection rate of a jet is 1/10 of the mass accretion onto the protostar.
The accretion rate of primordial star formation is expected to be considerably larger than that at present day, and it produces a stronger jet.
The life time of the jet also seems to be controlled by accretion in the present-day; a jet stops when mass accretion stops.
For Population III stars, the gas accretion does not halt within their lifetimes \citep{omukai01,omukai03}.
Therefore, a jet also may continue during the all lifetime of the protostar, and the strong jet propagates to disturb a surrounding medium significantly.
The disturbance of the medium could trigger the subsequent star formation as frequently observed in present-day star formation.

Assuming the power law growth of $B_0 \propto n_c^{2/3}$, the critical strength of the magnetic field, $B_0 = 10^{-9}$\,G at $n_c = 10^{3}\cm$ corresponds to $B_0 = 5\times 10^{-13}$\,G at $n_c = 0.01\cm$, which is much stronger than the background magnetic field of $10^{-18}$\,G derived by \citet{ichiki06}.
However, when the magnetic field is amplified to $B \sim 10^{-9}(\nc/10^3\cm)^{2/3}$\,G by some mechanisms, the magnetic field can affect the collapse of the primordial cloud.
% One possibility is a sheet collapse in which the sheet collapses in the direction perpendicular to the plane of the sheet.
% In this case, the parallel component of the magnetic with respective to the sheet is amplified as $B_\parallel \propto \rho$, exhibiting the significantly larger power law index than that of spherical collapse, 2/3.
% Therefore, the magnetic field is amplified from the background strength $10^{-18}$\,G to $10^{-13}$\,G in a range of $0.01\cm < \nc < 10^3 \cm$.
% Such a collapse is likely to occur in the case of extremely weak magnetic fields $\beta_{\rm p} \gg 1$.
Even if a cloud has a magnetic field weaker than the critical strength $B_0 = 10^{-9}$\,G, the magnetic field may play an important role after the protostar formation.
\citet{tan04} studies analytically the evolution of accretion disks around the first stars,  suggesting that magnetic fields amplified in the circumstellar disk eventually give rise to protostellar jets during the protostellar accretion phase.

Rotation promotes fragmentation when the first collapsed objects has the angular velocity of $\Omega_0 \gtrsim 10^{-17}(\nc/10^3\cm)^{2/3}$\,s$^{-1}$ as shown in Figure~\ref{fig:10}.
This condition coincides with that given by equation~(\ref{eq:fcri2}).
The fragmentation is expected to produces binary or multiple stellar system.
When a multiple stellar system is formed, some stars can be ejected by close encounters.
At the protostar formation epoch, the protostar has a mass of $M\simeq 10^{-3}\msun$.
The ejected proto-Population III stars may evolve to metal-free brown dwarfs or low-mass stars.
When a binary component in a multiple stellar system is ejected from the parent cloud by protostellar interaction, a low-mass metal free binary may also be appeared in the early universe. 
\citet{suda04} indicates that the extremely metal-poor ([Fe/H]$<$-5) stars \citep{christlieb01,frebel05} discovered recently are formed as binary members from metal-free gas, and then have been polluted by the companion stars during the stellar evolution.
\citet{komiya06} shows that a binary frequency in Population III star is comparable to or larger than that at present day.
In order to confirm the ejection scenario, the further long-term calculations are required.
%% as performed in the present-day star formation e.g., \citet{bate03}.

\acknowledgments
We thank K. Omukai for giving us the data of thermal evolution for primordial collapsing cloud.
We also thank T. Hanawa for contribution to the nested grid code.
We have greatly benefited from discussion with ~T. Tsuribe.
SI is grateful for the hospitality of KITP and interactions with the participants of the program ``Star Formation through Cosmic Time''.
Numerical computations were carried out on VPP5000 at Center for Computational Astrophysics, CfCA, of National Astronomical Observatory of Japan.
This work is supported by the Grants-in-Aid from MEXT (15740118, 16077202,18740113, 18740104).

%%%%%%%%%%%%%
%%% Table %%%
%%%%%%%%%%%%%
\begin{table}
\setlength{\tabcolsep}{7pt}
\caption{Models}
\label{table}
\scriptsize
\begin{center}
%%\scalebox{.5}{%
\rotatebox{90}{%
\begin{tabular}{c|ccccc|ccccccc} \hline
{\footnotesize Model} & $\beta_0$ & $\gamma_0$ & $\Omega_0$ {\scriptsize [s$^{-1}$]}& $B_0$ {\scriptsize [G]}  & $(M/\Phi)_{\rm norm}$ & 
$\Omega_{\rm ps}$ {\scriptsize [s$^{-1}$]} & $B_{\rm ps}$ {\scriptsize [G]} & 
$n_{\rm f}$ {\scriptsize ($\cm$)} & R$_{\rm s}$ {\scriptsize [AU]}&  $v_{\rm m}$ {\scriptsize [km\,s$^{-1}$]} & B/$\Omega$
\\ \hline
1     & 0.1 & 0                  & $7.2 \times 10^{-15}$  & 0          & $\infty$              &($6.3\times10^{-8}$) &(0)    &$9.6\times 10^{17}$ & 7.21  & ---  & $\Omega$ \\
2     & 0.1 & $2\times 10^{-9}$  & $7.2 \times 10^{-15}$  & $10^{-9}$  & $3\times 10^4$        &($4.0\times10^{-9}$) &(0.01) &$3.8\times 10^{17}$ & 69.8  & ---  & $\Omega$ \\
3     & 0.1 & $2\times 10^{-7}$  & $7.2 \times 10^{-15}$  & $10^{-8}$  & $3\times 10^3$        &($4.1\times10^{-8}$) &(0.04) &$9.1\times 10^{17}$ & 0.59  & ---  & $\Omega$ \\
4     & 0.1 & $2\times 10^{-5}$  & $7.2 \times 10^{-15}$  & $10^{-7}$  & $300$                 &($4.4\times10^{-8}$) &(0.3)  &$9.5\times 10^{17}$ & 2.63  & ---  & $\Omega$ \\
5     & 0.1 & $2\times 10^{-3}$  & $7.2 \times 10^{-15}$  & $10^{-6}$  & $30$                  &($9.2\times10^{-10}$)&(0.5)  &$5.3\times 10^{14}$ & 74.2  & ---  & $\Omega$ \\
6     & 0.1 & $2\times 10^{-1}$  & $7.2 \times 10^{-15}$  & $10^{-5}$  & $3$                   &($2.7\times10^{-10}$)&(0.2)  & ---                & ---   & ---  & --- \\
\hline
7     & 0.01 & 0                  & $2.3 \times 10^{-15}$  & 0          & $\infty$             &$6.9\times10^{-6}$& 0    &$1.0\times 10^{21}$  & 0.05 & ---  & $\Omega$ \\
8     & 0.01 & $2\times 10^{-9}$  & $2.3 \times 10^{-15}$  & $10^{-9}$  & $3\times 10^4$       &$7.8\times10^{-6}$& 1.6  &$9.9\times 10^{20}$  & 0.04 & ---  & $\Omega$ \\
9     & 0.01 & $2\times 10^{-7}$  & $2.3 \times 10^{-15}$  & $10^{-8}$  & $3\times 10^3$       &$7.6\times10^{-6}$& 21   &$1.0\times 10^{21}$  & 0.04 & ---  & $\Omega$ \\
10    & 0.01 & $2\times 10^{-5}$  & $2.3 \times 10^{-15}$  & $10^{-7}$  & $300$                &$7.9\times10^{-6}$& 213  &$1.1\times 10^{21}$  & 0.03 & ---  & $\Omega$ \\
11    & 0.01 & $2\times 10^{-3}$  & $2.3 \times 10^{-15}$  & $10^{-6}$  & $30$                 &$1.1\times10^{-6}$& 2471 &$(1.2\times 10^{21})$& ---  & 37.9 & $\Omega$ \\
12    & 0.01 & $2\times 10^{-1}$  & $2.3 \times 10^{-15}$  & $10^{-5}$  & $3$                  &$5.6\times10^{-7}$& 0.01 & ---                 & ---  & 55.9 & $B$ \\
\hline
13     & $10^{-3}$ & 0                  & $7.2 \times 10^{-16}$  & 0          & $\infty$       &($5.9\times10^{-7}$)&(0)    &$4.0\times 10^{18}$  & 0.19 & --- & $\Omega$ \\
14     & $10^{-3}$ & $2\times 10^{-9}$  & $7.2 \times 10^{-16}$  & $10^{-9}$  & $3\times 10^4$ &($6.4\times10^{-7}$)&(0.2)  &$4.8\times 10^{18}$  & 0.16 & --- & $\Omega$ \\
15     & $10^{-3}$ & $2\times 10^{-7}$  & $7.2 \times 10^{-16}$  & $10^{-8}$  & $3\times 10^3$ &($6.1\times10^{-7}$)&(1.7)  &$4.7\times 10^{18}$  & 0.17 & --- & $\Omega$ \\
16     & $10^{-3}$ & $2\times 10^{-5}$  & $7.2 \times 10^{-16}$  & $10^{-7}$  & $300$          &($6.8\times10^{-7}$)&(18.5) &$5.4\times 10^{18}$  & 0.08 & --- & $\Omega$ \\
17     & $10^{-3}$ & $2\times 10^{-3}$  & $7.2 \times 10^{-16}$  & $10^{-6}$  & $30$           &($2.3\times10^{-5}$)&(6659) &($2.7\times 10^{21}$)& ---  & 32.5& $B$ \\
18     & $10^{-3}$ & $2\times 10^{-1}$  & $7.2 \times 10^{-16}$  & $10^{-5}$  & $3$            &($1.5\times10^{-5}$)&(35590)& -- -                & ---  & 84.6& $B$ \\
\hline
19     & $10^{-4}$ & 0                  & $2.3 \times 10^{-16}$  & 0          & $\infty$       &($1.6\times10^{-6}$)&(0)   &$2.4\times 10^{19}$  & 0.08 & ---  & $\Omega$ \\
20     & $10^{-4}$ & $2\times 10^{-9}$  & $2.3 \times 10^{-16}$  & $10^{-9}$  & $3\times 10^4$ &($6.5\times10^{-7}$)&(0.6) &$7.9\times 10^{18}$  & 0.29 & ---  & $\Omega$ \\
21     & $10^{-4}$ & $2\times 10^{-7}$  & $2.3 \times 10^{-16}$  & $10^{-8}$  & $3\times 10^3$ &($6.7\times10^{-7}$)&(5.9) &$8.4\times 10^{18}$  & 0.11 & ---  & $\Omega$ \\
22     & $10^{-4}$ & $2\times 10^{-5}$  & $2.3 \times 10^{-16}$  & $10^{-7}$  & $300$          &($9.7\times10^{-7}$)&(82)  &$1.6\times 10^{19}$  & 0.08 & ---  & $\Omega$ \\
23     & $10^{-4}$ & $2\times 10^{-3}$  & $2.3 \times 10^{-16}$  & $10^{-6}$  & $30$           &$8.1\times10^{-6}$& 13631  &---                  & ---  & 66.3 & $B$ \\
24     & $10^{-4}$ & $2\times 10^{-1}$  & $2.3 \times 10^{-16}$  & $10^{-5}$  & $3$            &$3.5\times10^{-6}$& 31935  &---                  & ---  & 46.8 & $B$ \\
\hline
25     & $10^{-5}$ & 0                  & $7.2 \times 10^{-17}$  & 0          & $\infty$       &$1.1\times10^{-5}$& 0     &$1.3\times 10^{21}$  & 0.07 & ---  & $\Omega$ \\
26     & $10^{-5}$ & $2\times 10^{-9}$  & $7.2 \times 10^{-17}$  & $10^{-9}$  & $3\times 10^4$ &$1.0\times10^{-5}$& 25    &$1.0\times 10^{21}$  & 0.11 & ---  & $\Omega$ \\
27     & $10^{-5}$ & $2\times 10^{-7}$  & $7.2 \times 10^{-17}$  & $10^{-8}$  & $3\times 10^3$ &$1.0\times10^{-5}$& 238   &$8.9\times 10^{20}$  & 0.03 & ---  & $\Omega$ \\
28     & $10^{-5}$ & $2\times 10^{-5}$  & $7.2 \times 10^{-17}$  & $10^{-7}$  & $300$          &$1.1\times10^{-5}$& 4251  &($1.8\times 10^{21}$)& ---  & 16.9 & $B$ \\
29     & $10^{-5}$ & $2\times 10^{-3}$  & $7.2 \times 10^{-17}$  & $10^{-6}$  & $30$           &$1.0\times10^{-5}$& 20634 & ---                 & ---  & 79.3 & $B$ \\
30     & $10^{-5}$ & $2\times 10^{-1}$  & $7.2 \times 10^{-17}$  & $10^{-5}$  & $3$            &$1.1\times10^{-6}$& 30545 & ---                 & ---  & 75.4 & $B$ \\
\hline
31     & $10^{-6}$ & 0                  & $2.3 \times 10^{-17}$  & 0          & $\infty$       &$1.0\times10^{-5}$& 0     & ---                 & ---  & ---  & $\Omega$ \\
32     & $10^{-6}$ & $2\times 10^{-9}$  & $2.3 \times 10^{-17}$  & $10^{-9}$  & $3\times 10^4$ &$1.1\times10^{-5}$& 161   & ---                 & ---  & ---  & $\Omega$ \\
33     & $10^{-6}$ & $2\times 10^{-7}$  & $2.3 \times 10^{-17}$  & $10^{-8}$  & $3\times 10^3$ &$1.1\times10^{-5}$& 16371 & ---                 & ---  & 34.5  & $\Omega$ \\
34     & $10^{-6}$ & $2\times 10^{-5}$  & $2.3 \times 10^{-17}$  & $10^{-7}$  & $300$          &$7.7\times10^{-6}$& 11616 & ---                 & ---  & 50.7 & $B$ \\
35     & $10^{-6}$ & $2\times 10^{-3}$  & $2.3 \times 10^{-17}$  & $10^{-6}$  & $30$           &$3.7\times10^{-6}$& 25890 & ---                 & ---  & 35.1 & $B$ \\
36     & $10^{-6}$ & $2\times 10^{-1}$  & $2.3 \times 10^{-17}$  & $10^{-5}$  & $3$            &$3.3\times10^{-7}$& 29874 & ---                 & ---  & 67.6 & $B$ \\
\hline
\end{tabular}}
\end{center}
\end{table}

\clearpage
%%%%%%%%%%%%%%%%%%%%
%%%%% FIGURES %%%%%%
%%%%%%%%%%%%%%%%%%%%

%%%%%%%%%%%%
%% Fig. 1 %%
%%%%%%%%%%%%
%%%%%%%%%%%%
%% Fig. 2 %%
%%%%%%%%%%%%
%%%%%%%%%%%%
%% Fig. 3 %%
%%%%%%%%%%%%
%%%%%%%%%%%%
%% Fig. 4 %%
%%%%%%%%%%%%
%%%%%%%%%%%%
%% Fig. 5 %%
%%%%%%%%%%%%
%%%%%%%%%%%%
%% Fig. 6 %%
%%%%%%%%%%%%
%%%%%%%%%%%%
%% Fig. 7 %%
%%%%%%%%%%%%
%%%%%%%%%%%%
%% Fig. 8 %%
%%%%%%%%%%%%
%%%%%%%%%%%%
%% Fig. 9 %%
%%%%%%%%%%%%
%%%%%%%%%%%%
%% Fig. 10%%
%%%%%%%%%%%%
\end{document}